\documentclass[trackchanges,twocolumn]{aastex701}

\received{xxxx, 2026}
\revised{xxxx, 2026}
\accepted{xxxx, 2026}
\submitjournal{ApJ}
\newcommand{\rev}[1]{}

\usepackage{multirow}
\usepackage{comment}
\usepackage{subfigure}
\usepackage{amsmath}
\usepackage{booktabs}
\usepackage{CJK}

\shorttitle{Multi-timescale radio variability of 4C+72.28}
\shortauthors{Wang, Mufakharov, Tan, et al.}
\graphicspath{{./}}

\begin{document}
\begin{CJK*}{UTF8}{gbsn}
\title{SMMAN Studies of AGN Variability. I. Long-term Spectral Evolution and a Candidate \(\sim1000\)-day Quasi-periodic Modulation in 4C+72.28}

\correspondingauthor{Lang Cui}
\email{cuilang@xao.ac.cn}

\author[0000-0001-8221-9601]{Xin Wang (王鑫)}
\affiliation{State Key Laboratory of Radio Astronomy and Technology, Xinjiang Astronomical Observatory, CAS, 150 Science 1-Street, Urumqi 830011, China}
\affiliation{College of Astronomy and Space Science, University of Chinese Academy of Sciences, No.1 Yanqihu East Road, Beijing 101408, China}
\email{wangxin2019@xao.ac.cn}

\author[0000-0001-9984-127X]{Timur Mufakharov}
\affiliation{State Key Laboratory of Radio Astronomy and Technology, Xinjiang Astronomical Observatory, CAS, 150 Science 1-Street, Urumqi 830011, China}
\affiliation{Special Astrophysical Observatory of the Russian Academy of Sciences, Nizhny Arkhyz, 369167, Russia}
\email{timur.mufakharov@gmail.com}

\author[0009-0006-7762-1413]{Guiping Tan (谭桂萍)}
\affiliation{State Key Laboratory of Radio Astronomy and Technology, Xinjiang Astronomical Observatory, CAS, 150 Science 1-Street, Urumqi 830011, China}
\email{tanguiping@xao.ac.cn}

\author[0009-0007-7436-0673]{Zhenhua Han(韩振华)}
\affiliation{School of Electronic and Communication Engineering, Kashi University, Kashi 844000, China}
\email[]{hzh202312@163.com}

\author[0000-0002-8684-7303]{Ning Chang (常宁)}
\affiliation{State Key Laboratory of Radio Astronomy and Technology, Xinjiang Astronomical Observatory, CAS, 150 Science 1-Street, Urumqi 830011, China}
\email{changning@xao.ac.cn}

\author[0000-0003-4671-1740]{Qi Yuan (袁琦)}
\affiliation{Changchun Observatory, National Astronomical Observatories, Chinese Academy of Sciences, Changchun 130117, People’s Republic of China}
\email{yuanqi@cho.ac.cn}

\author[0000-0001-8221-9601]{Yunzhuo Wang (王运卓)}
\affiliation{State Key Laboratory of Radio Astronomy and Technology, Xinjiang Astronomical Observatory, CAS, 150 Science 1-Street, Urumqi 830011, China}
\affiliation{College of Astronomy and Space Science, University of Chinese Academy of Sciences, No.1 Yanqihu East Road, Beijing 101408, China}
\email{wangyunzhuo@xao.ac.cn}

\author[0000-0001-9815-2579]{Xiang Liu (刘祥)}
\affiliation{State Key Laboratory of Radio Astronomy and Technology, Xinjiang Astronomical Observatory, CAS, 150 Science 1-Street, Urumqi 830011, China}
\affiliation{College of Astronomy and Space Science, University of Chinese Academy of Sciences, No.1 Yanqihu East Road, Beijing 101408, China}
\affiliation{Xinjiang Key Laboratory of Radio Astrophysics, 150 Science 1-Street, Urumqi, Xinjiang, 830011, China}
\email{liux@xao.ac.cn}


\author[0000-0003-0721-5509]{Lang Cui (崔朗)}
\affiliation{State Key Laboratory of Radio Astronomy and Technology, Xinjiang Astronomical Observatory, CAS, 150 Science 1-Street, Urumqi 830011, China}
\affiliation{College of Astronomy and Space Science, University of Chinese Academy of Sciences, No.1 Yanqihu East Road, Beijing 101408, China}
\affiliation{Xinjiang Key Laboratory of Radio Astrophysics, 150 Science 1-Street, Urumqi, Xinjiang, 830011, China}
\email{cuilang@xao.ac.cn}

\begin{abstract}
The SMMAN project has carried out nine years of radio flux monitoring in the C and K bands of the blazar 4C+72.28. The two bands show markedly different secular evolution, with a substantially stronger K-band decline producing a gradual decrease in the C--K spectral index. 
After accounting for this secular evolution, multiple timing diagnostics reveal a candidate quasi-periodic modulation with a characteristic timescale of $\sim1000$\,d, most clearly in the C band. 
The C-band GLSP peak reaches a local significance of $3.29\sigma$, while its global significance remains modest at $\sim2\sigma$ and the monitoring baseline covers only about three cycles.
Independent and joint MCMC fits support a phenomenological decomposition for both bands into a long-term linear trend and a recurrent component. Under the shared-period assumption, the joint fit returns a characteristic timescale of approximately $1027$\,d and similar fractional recurrent amplitudes of $\simeq5.45\%$. Information-criterion comparisons favor the combined linear-plus-sinusoidal model in both bands.
Cross-correlation analysis shows a nominal K-leading offset in the original light curves but a centroid consistent with zero lag after detrending, although the lag distributions overlap substantially. 
These results are consistent with a common recurrent modulation superposed on slower frequency-dependent jet evolution. Geometrical Doppler-factor modulation provides one plausible interpretation, although intrinsic and opacity-related processes cannot be excluded.

\end{abstract}

\keywords{\uat{Active galactic nuclei}{16} --- \uat{Blazars}{164} --- \uat{BL Lacertae objects}{158} ---\uat{Radio continuum emission}{1340} --- \uat{Time domain astronomy}{2109}}

\section{Introduction} \label{sec:intro}
Blazars are radio-loud active galactic nuclei (AGNs) whose relativistic jets are oriented close to the observer's line of sight. Their broadband emission is dominated by Doppler-boosted non-thermal radiation from the jet \citep{1979ApJ...232...34B,1995PASP..107..803U,2019ARA&A..57..467B}. Flux variability is observed across the electromagnetic spectrum and can be produced by changes in particle injection and acceleration, radiative cooling, magnetic fields, opacity, jet geometry, and the accretion flow. At centimeter and millimeter wavelengths, long-term monitoring probes the evolution of synchrotron-emitting regions on different scales along the jet and can reveal propagating disturbances, secular spectral evolution, and geometrical changes in the emitting flow \citep{1985ApJ...298..114M,2007A&A...469..899H}. Long-term radio light curves therefore provide an important means of studying the connection between the central engine and the parsec-scale jet.

Quasi-periodic oscillations (QPOs) and quasi-periodic variations have been reported in the light curves of a number of AGNs. Proposed mechanisms include periodic or quasi-periodic modulation of the accretion flow, Lense--Thirring precession, binary-induced orbital or precessional motion, helical trajectories of emitting features, and periodic changes in the jet viewing angle and Doppler factor \citep{1918PhyZ...19..156L,1998ApJ...492L..59S,1999A&A...347...30V}. However, AGN variability is generally stochastic and frequently exhibits a red-noise power spectrum \citep{2005A&A...431..391V}. Irregular sampling, finite observational baselines, window-function effects, and searches over multiple frequencies can produce apparently significant periodogram peaks. Consequently, QPO claims require careful treatment of the underlying stochastic process and of the look-elsewhere effect \citep{2008MNRAS.385.1279B,2013MNRAS.433..907E,2018ApJS..236...16V,2020A&A...634A.120A}. The consistency of a candidate timescale among frequency-domain, time--frequency, and time-domain methods is valuable, but does not by itself make these tests statistically independent. Local and global significance estimates against an appropriate red-noise null hypothesis therefore remain essential.

Several well-known long-timescale candidates have been reported outside the radio band. The approximately 12-yr optical outburst cycle of OJ\,287 has motivated binary-supermassive-black-hole interpretations \citep{1988ApJ...325..628S}, while an approximately 2-yr modulation in the $\gamma$-ray and multiwavelength emission of PG\,1553+113 has been discussed in terms of recurrent activity or geometrical modulation \citep{2015ApJ...813L..41A}. Radio monitoring provides a complementary view because it can cover several decades and trace the evolution of synchrotron-emitting jet components at progressively larger distances from the central engine.

Candidate radio QPOs have been reported over timescales ranging from several months to several years. Examples include the 120--150 d modulation in the 15-GHz light curve of J1359+4011 \citep{2013MNRAS.436L.114K}, year-scale harmonic components in 1156+295 \citep{2014MNRAS.443...58W}, a $270\pm26$ d feature in PKS\,0219$-$164 \citep{2017ApJ...847....7B}, and a $965\pm50$ d signal found at 4.8, 8.0, and 14.5 GHz in AO\,0235+164 \citep{2021MNRAS.501.5997T}. Multiple year-scale candidates have also been reported in OT\,081 \citep{2021MNRAS.506.1540L}, while a possible 4.69-yr modulation was identified in the 15-GHz light curve of PKS\,J2134$-$0153 \citep{2021MNRAS.506.3791R}. One of the most notable radio cases is PKS\,2131$-$021, whose 45-yr light curve contains two separated episodes of nearly sinusoidal modulation with mutually consistent periods and phases \citep{2022ApJ...926L..35O}. More recent examples include a possible $352\pm23$ d modulation in the 37-GHz light curve of S5\,0716+714 \citep{2023ApJ...943..157L}, an approximately 611 d signal in PKS\,J2156$-$0037 \citep{2024MNRAS.531.3927M}, and an approximately 2.4-yr candidate in the 37-GHz light curve of Ton\,599 \citep{2026A&A...707A.371G}. Together, these studies demonstrate the potential of long-term radio monitoring for identifying recurrent jet activity, but also show that the inferred significance and persistence of a candidate can depend on the observational baseline, the searched frequency interval, and the adopted stochastic-noise model.

The source 4C+72.28, also known as S5\,2010+723 and J2009+7229, has long been included in bright radio-source and radio-selected BL~Lac samples \citep{1981A&AS...45..367K,1990AJ.....99....1K, 1995MNRAS.277.1477P}. It is classified as a BL~Lac object, but its redshift remains undetermined. Multifrequency studies classify it as a low-synchrotron-peaked source and show that it is generally faint at optical wavelengths \citep{2010A&A...512A..74M}. Multi-epoch 5-GHz observations from the Caltech--Jodrell Bank flat-spectrum survey provide parsec-scale VLBI images and model fits of its compact core--jet structure \citep{2007A&A...472..763B,2008A&A...484..119B}. The source has also been included in high-frequency radio and microwave surveys \citep{2011MNRAS.410.2690P,2013A&A...553A.107C} and in the RATAN-600 multifrequency monitoring sample of BL~Lac objects \citep{2017AN....338..700M}. Nevertheless, these observations are sparse or were analysed only as part of larger samples, and no dedicated study of its long-term radio periodicity has previously been presented. 

To our knowledge, this is the first dedicated study of the long-term radio variability of 4C+72.28, including its secular spectral evolution and possible recurrent modulation. We use the nearly decade-long 4.8- and 23.6-GHz light curves obtained by the quasi-Simultaneous Multi-frequency Monitoring of AGNs with the Nanshan 26-m Radio Telescope (SMMAN) project \citep{2026ApJS..283...69C} to investigate variability on multiple timescales. We first characterize the frequency-dependent secular evolution and the corresponding C--K spectral-index changes in Section~\ref{sec:lse}. 
After separating the long-term component, we investigate whether the long-term radio variability of 4C+72.28 contains a quasi-periodic component on timescales substantially shorter than the observed secular evolution, and whether such a component is present consistently across radio frequencies in Section~\ref{sec:qpo}.
We then use independent and joint MCMC fits and information-criterion comparisons to test a phenomenological decomposition into secular and recurrent components \rev{in Section~\ref{sec:mcmc}}. 
Rather than treating a periodic signal as an isolated phenomenon, we examine it together with the long-term frequency-dependent evolution of the radio spectrum via cross-correlation analysis in Section~\ref{sec:ccf}.
Finally, we discuss whether the observed behavior can be understood in terms of a common geometrical modulation superposed on slower intrinsic and/or opacity-related evolution of the radio-emitting jet in Section~\ref{sec:disc}.

\section{Observation} \label{sec:obs}

\begin{figure*}
    \centering
    \includegraphics[angle=0,scale=0.65]{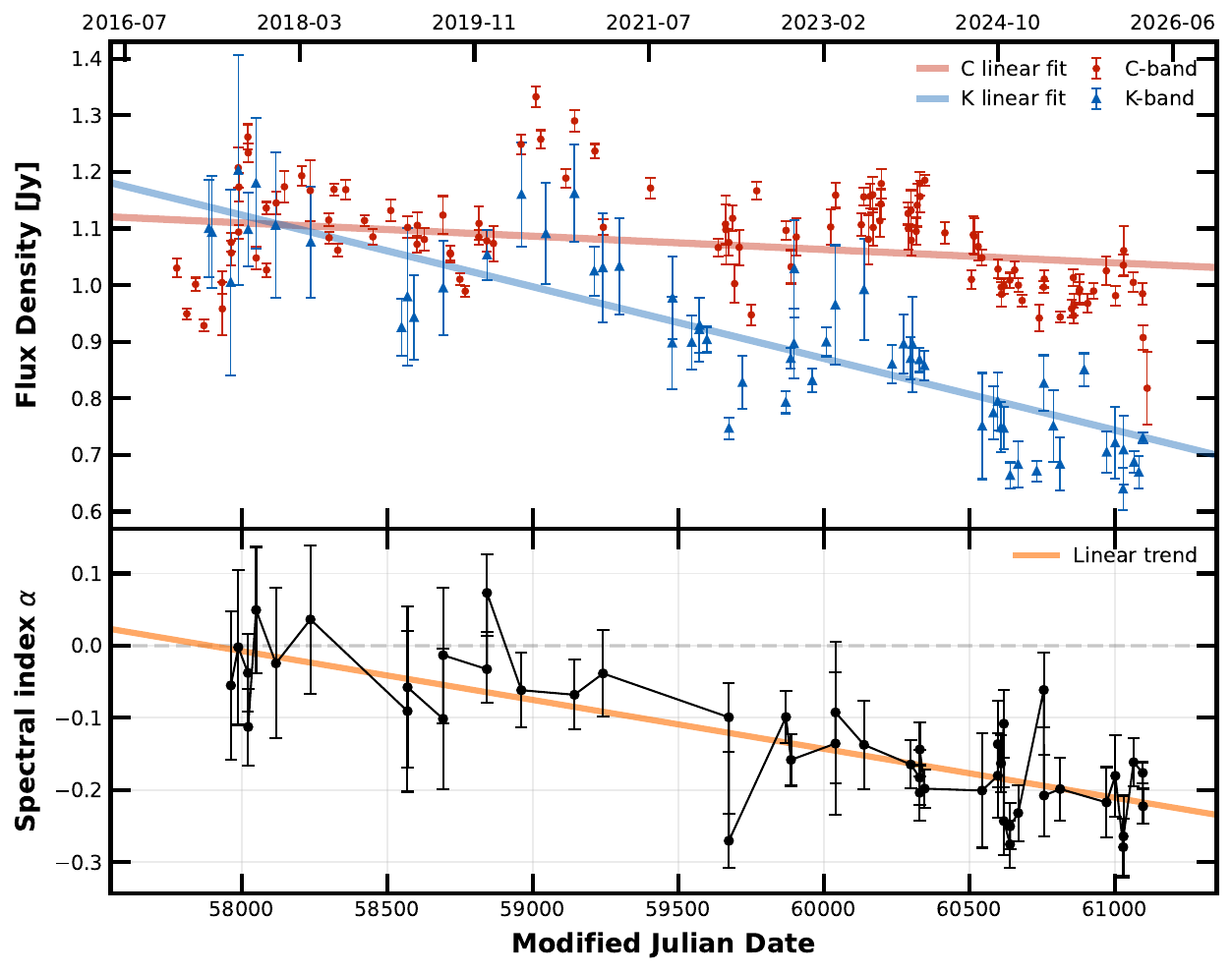}
    \caption{1-day binned C- and K-band light curves of 4C+72.28. In the top panel, the red circles and blue triangles represent the C-band and K-band flux densities, respectively. The red and blue solid lines show the corresponding linear trends obtained by least-squares fitting to the original, unbinned light-curves. In the bottom panel, the spectral index \(\alpha\) vs. MJD is shown. The solid line represents the linear fit to the spectral index. The gray dashed line marks \(\alpha=0\), and the error bars indicate the estimation uncertainties.}
    \label{fig:lcs}
\end{figure*}

This study uses the long-term monitoring data from the SMMAN project, which is a radio flux density monitoring program for typical AGNs with the Nanshan 26-m Radio Telescope (NSRT) of Xinjiang Astronomical Observatory \citep{2026ApJS..283...69C}. The SMMAN project started in 2017 and is still ongoing. The nominal observing interval is from \rev{a week} to about one month, although longer gaps exist because of weather, scheduling, and data quality limitations. In this work, we use the available radio monitoring data at C band and K band. The central frequencies are 4.8 GHz for C band and 23.6 GHz for K band. Therefore, the SMMAN project provides a long-term radio flux density data set of the blazar 4C+72.28 from 2017 to 2026.


\rev{4C+72.28 was chosen for this dedicated analysis on the basis of its unusual recurrent variability pattern. We did not perform a systematic search for QPOs over the entire sample; therefore, the reported false-alarm probabilities do not include a sample-wide trials correction.}

The SMMAN monitoring observations are performed with the cross-scan mode. In this mode, the target source is scanned along two orthogonal directions, azimuth and elevation. The standard data reduction procedure includes signal extraction, Gaussian fitting, pointing calibration, atmospheric opacity calibration, elevation gain calibration, and absolute flux density calibration. The final flux density scale of the monitored sources is mainly determined by three calibrators, 3C 48, NGC 7027, and 3C 286. The flux density uncertainties are taken from the SMMAN reduction pipeline. They include the fitting uncertainty and the relative calibration uncertainty, which is mainly estimated from the scatter of the calibrator measurements.

For the C-band light curve, the time coverage is from MJD 57777 to MJD 61111 (i.e., January 24, 2017, to March 12, 2026). The original light curve contains 203 valid measurements. To show the light curve more clearly, we combine the data points observed within the same day by using a 1-day binning procedure. After this binning, 115 binned epochs remain. The median observing cadence, calculated from the adjacent binned epochs, is about 18.6 days.

For the K-band light curve, the time coverage is from MJD 57887 to MJD 61096 (i.e., May 14, 2017, to February 25, 2026). The original light curve contains 133 valid measurements. After the same 1-day binning procedure, 60 binned epochs remain, and the median observing interval is about 31.1 days. The 1-day binned C- and K-band light curves of 4C+72.28 are shown in Figure \ref{fig:lcs}. It should be noted that the 1-day binned data are only used for display in the figure.

\rev{The 203 C-band and 133 K-band measurements used here are all valid data from the SMMAN without subjectively selected within specific intervals. The only quality-control criteria which flagged data points deviating more than five times the RMS of the mean flux density is applied; no other selection or rejection criteria were applied.}

\section{Long-term chromatic variability} \label{sec:lse}

By visual inspection, the K-band light curve shows an obvious long-term linear trend. Using a least-squares linear fit, the light curves show slopes of $-2.36\times10^{-5}\ {\rm Jy\,d^{-1}}$ and $-1.27\times10^{-4}\ {\rm Jy\,d^{-1}}$ in the C and K bands, respectively. Based on the fitted slopes and the corresponding observational time spans of about 3330 d, the linear models give total flux-density changes of approximately $-0.08$ Jy and $-0.41$ Jy in the C and K bands, indicating a more obvious variation in the K band than in the C band.

A frequency-dependent secular variability suggests that there could also be a secular evolution of the spectral index. To investigate the long-term spectral change, time bins with a maximum duration of 2 days were used, and the binned quasi-simultaneous C-K spectral index was calculated as
\begin{equation}
\alpha =
\frac{\ln (f_{K}/ f_{C})}
{\ln (\nu_{K}/\nu_{C})},
\end{equation}
where \(f_{C}\) and \(f_{K}\) are the flux densities in the C and K bands, respectively, and \(\nu_{C}\) and \(\nu_{K}\) are their corresponding effective frequencies. The uncertainty of the spectral index was estimated as 
\begin{equation}
\sigma_\alpha =
\frac{1}
{\ln (\nu_{K}/\nu_{C})}\sqrt{(\frac{\sigma_{\rm c}}{f_{\rm c}})^2+(\frac{\sigma_{\rm k}}{f_{\rm k}})^2},
\end{equation}
where \(\sigma_{\rm c}\) and \(\sigma_{\rm k}\) denote the measurement errors in the C- and K-band.
The temporal evolution of the spectral index is shown in Figure \ref{fig:lcs}. We performed a simple linear fit to the spectral-index measurements, with the best-fitting relation indicated by the red line. The best-fit slope, corresponding to the spectral evolution rate, is \(-6.78\times 10^{-5}\ {\rm d}^{-1}\), giving total index change of \(\sim0.226\) decrease over the entire time span. The chromatic secular evolution suggests that the observed radio variability cannot be described by a single achromatic component, motivating further separation of long-term and recurrent variability.

\section{Search for quasi-periodicity} \label{sec:qpo}
To isolate possible quasi-periodic components and reduce the red-noise background, we remove the significant linear trend in the K-band light curve, while the C-band data are left unchanged, because its secular evolution is comparatively weak. 
\rev{This treatment serves to meet the stationarity requirement of the analysis algorithm for time series.}

We characterize the quasi-periodicity using three complementary approaches: the generalized Lomb--Scargle periodogram (GLSP), the weighted wavelet Z-transform (WWZ), and the auto-correlation function estimated with the z-transformed discrete correlation function (Auto-ZDCF). The GLSP provides a frequency-domain characterization of the dominant timescale, while the WWZ examines whether the corresponding variability persists over the observational span. The Auto-ZDCF provides a methodologically distinct time-domain consistency check.  

We evaluated the statistical significance of the  {candidate} signal against stochastic red-noise plus measurement white-noise light curves following the Monte Carlo framework of \citet{2013MNRAS.433..907E}. The stochastic background was constructed from a broken power-law power spectral density with an independently constrained white-noise level. We generated $10^5$ realizations for each band and repeated the same GLSP, WWZ, and Auto-ZDCF analyses on the simulated light curves. The detailed construction of the stochastic background and the calculation of local and global significance levels are described in Appendices~\ref{appendix: rdn}/\ref{appendix: sig}.  {Additional robustness tests of the detrending procedure and of the stochastic background modeling are presented in Appendix \ref{appendix:robustness}.}

\subsection{GLSP}

\begin{figure*}[htbp]
    \centering
    \begin{minipage}[b]{0.48\textwidth}
        \centering
        \includegraphics[width=\textwidth]{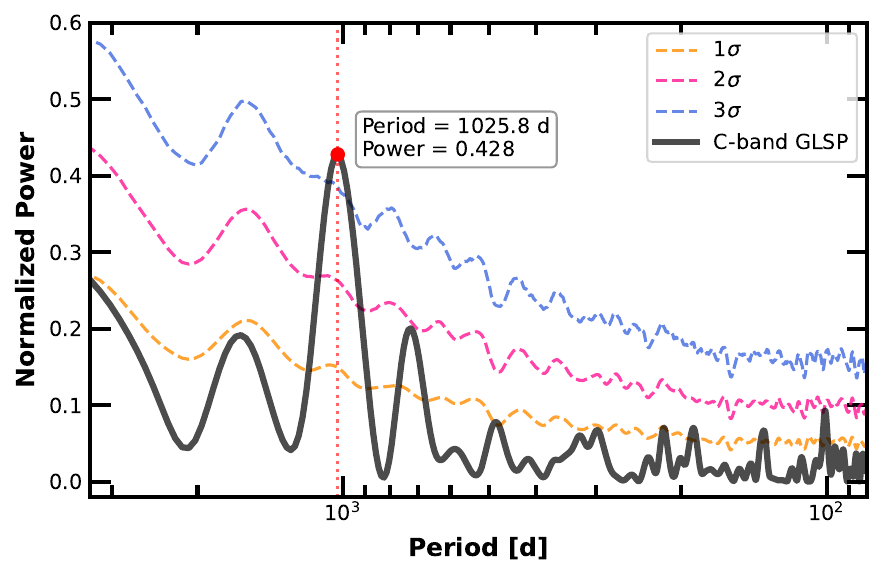}
        \vspace{0.5em} 
        \textbf{(a)}
    \end{minipage}
    \hfill 
    \begin{minipage}[b]{0.48\textwidth}
        \centering
        \includegraphics[width=\textwidth]{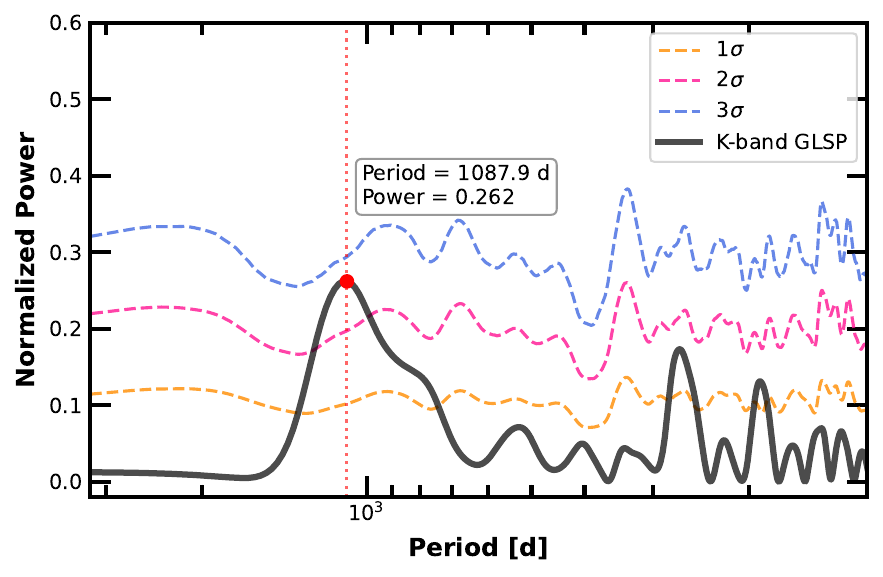}
        \vspace{0.5em}
        \textbf{(b)}
    \end{minipage}
    
    \caption{Generalized Lomb-Scargle Periodograms of 4C+72.28 in C- (a) and detrended K-band (b). 
    The orange, pink and blue dashed lines indicate the $1\sigma$, $2\sigma$, and $3\sigma$ local significance levels. The red point and dashed vertical line mark the strongest candidate peaks at $1025.76$~d and $1087.81$~d for the C and K bands, respectively.}
    \label{fig:glsp}
\end{figure*}

One of the most widely used methods for periodicity analysis is the frequency-domain power spectrum, or periodogram, based on the Fourier transform. However, astronomical observations, especially long-term monitoring data such as the SMMAN dataset used in this study, are often unevenly sampled. In such data, directly applying the conventional discrete Fourier transform may cause the estimated power spectrum to be strongly affected by the sampling window function. This may produce spectral leakage, aliasing, or spurious peaks \citep{1975Ap&SS..36..137D, 2018ApJS..236...16V}. To analyze periodic signals in irregularly sampled time series, \citet{1976Ap&SS..39..447L} and \citet{1982ApJ...263..835S} developed the Lomb--Scargle periodogram (LSP), which provides a more robust estimate of spectral power by fitting sinusoidal models to unevenly sampled data. Based on this framework, \citet{2009A&A...496..577Z} further introduced data weights associated with measurement uncertainties and developed the generalized Lomb--Scargle periodogram (GLSP).

For each band, the minimum frequency was set to $f_{\min}=1/T$, where $T=t_{\max}-t_{\min}$ is the baseline of the corresponding light curve. Thus, the maximum searched period was $P_{\max}\equiv T$. The maximum frequency was set to $f_{\max}=1/(5\cdot\bar{\tau})$, where $\bar{\tau}$ is the mean sample interval. The corresponding minimum searched period was therefore $P_{\min}=5\cdot\bar{\tau}$. \rev{Because Nyquist frequency cannot be uniquely defined for irregularly sampled data}, this relatively conservative frequency limit helps reduce the possible influence of aliasing in the high frequency range. Because the target signals in this study are located in a relatively low-frequency range, we adopted an oversampled frequency spacing of $\Delta f=1/(20\cdot T)$. \rev{This oversampling only increases the numerical sampling density, and reduces the possibility that a sparse frequency grid misses the local maxima in the periodogram, but does not improve the intrinsic frequency resolution.}

The resulting periodograms are shown in Figure~\ref{fig:glsp}. The C-band periodogram exhibits a prominent peak at $P_{\rm C}=1025.76$~d, with a normalized power of 0.428. The corresponding local significance is $3.29\sigma$, whereas the global significance over the band-specific search interval is $1.95\sigma$. Thus, the peak is prominent relative to the adopted stochastic background at its particular frequency, but the global significance is insufficient to establish a definitive QPO detection. The detrended K-band periodogram shows a broader and weaker peak at $P_{\rm K}=1087.81$~d, with a normalized power of 0.262. Its local significance is $2.69\sigma$, while the global significance is $2.18\sigma$. The weaker K-band response is consistent with the larger measurement and systematic uncertainties of the K-band monitoring data. Nevertheless, the characteristic timescale is broadly consistent with that identified in the C band.

To estimate the influence of measurement uncertainties on the localization of the characteristic timescale, we further performed $10^4$ flux randomization (FR) realizations for each light curve. \rev{In this procedure, the observing epochs are held fixed, and the flux densities are drawn from Gaussian distributions centered on the observed values with standard deviations equal to the measurement uncertainties.} For each realization, the GLSP was calculated on the same frequency grid and the peak was recorded. The resulting peak-period distributions were unimodal. We characterized these distributions with a Gaussian fit and adopted its centroid and standard deviation as an empirical estimate of the timescale and its measurement-error propagation, respectively. The resulting values are $P_{\rm C}=1027.97\pm3.31$~d and $P_{\rm K}=1099.35\pm25.30$~d. The KDE modes, $1028.11$~d and $1095.18$~d for the C and K bands, respectively, are consistent with the Gaussian-fit estimates. The uncertainties obtained from the FR distributions only quantify the propagation of the measurement errors. Thus they should not be interpreted as the full uncertainty associated with the finite observational baseline, sampling window, or the method. The observed GLSP peak periods are therefore retained as the primary statistics for the significance analysis, while the FR results provide an estimation of the localization uncertainty.

\subsection{Time-Frequency and Time-Domain Consistency} \label{sec:wwz-acf}

\begin{figure*}[htbp]
    \centering
    \begin{minipage}[b]{0.48\textwidth}
        \centering
        \includegraphics[width=\textwidth]{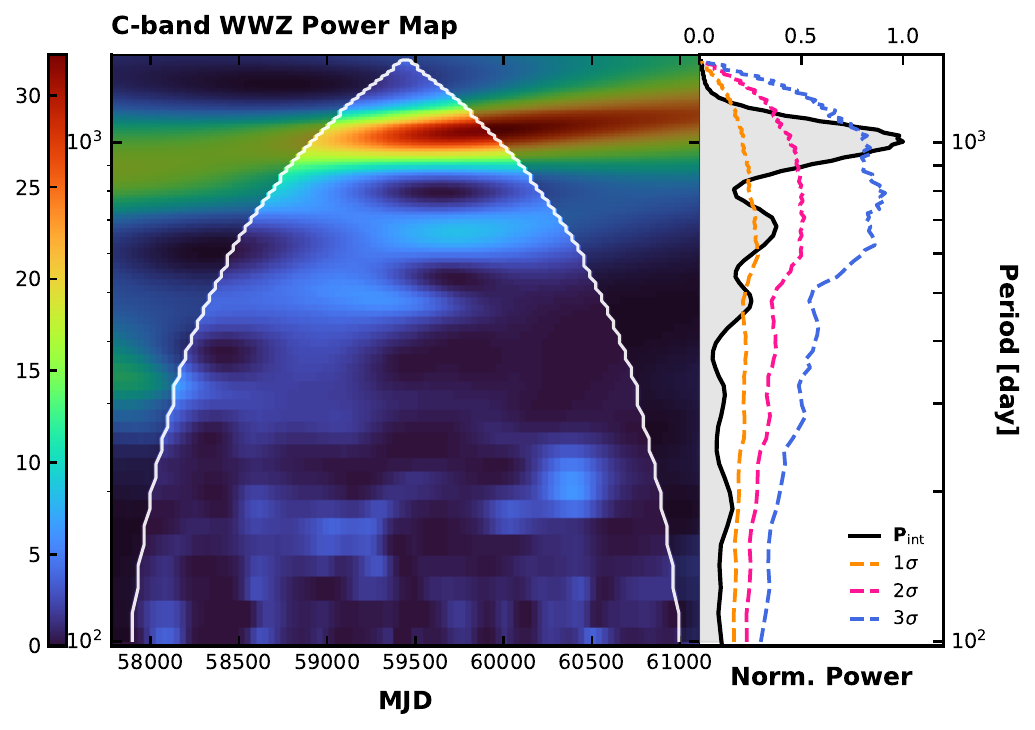}
        \vspace{0.5em} 
        \textbf{(a)}
    \end{minipage}
    \hfill 
    \begin{minipage}[b]{0.48\textwidth}
        \centering
        \includegraphics[width=\textwidth]{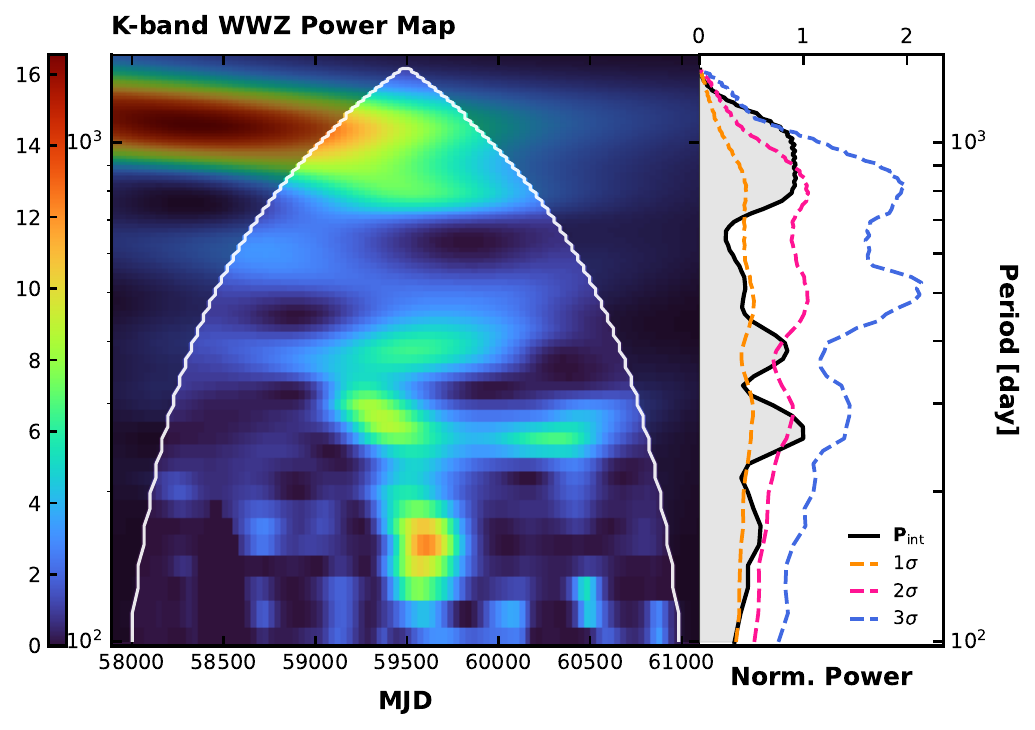}
        \vspace{0.5em}
        \textbf{(b)}
    \end{minipage}
    
    \caption{WWZ power map and integrated power spectra of 4C+72.28 in the C-band (a) and K-band (b). For each figure, the left main panel shows the distribution of WWZ power in the time--frequency domain, with the power intensity indicated by the color bar on the left. The shaded region outside the white curve denotes the area affected by edge effects, which is the COI, while the region inside the curve corresponds to reliable signals. The right panel presents the normalized power spectra obtained by integrating the WWZ power over time, excluding the COI, with the $1\sigma$, $2\sigma$, and $3\sigma$ local significance levels marked.}
    \label{fig:wwz}
\end{figure*}

Wavelet analysis is a classical time-frequency technique. \cite{1996AJ....112.1709F} developed the weighted wavelet Z-transform (WWZ) method specifically for unevenly sampled astronomical time series. Compared with conventional power-spectrum analyses, the WWZ not only characterizes the power distribution in the frequency domain, but also reveals the temporal evolution of different periodic components. It is therefore well suited for the analysis of non-stationary signals and provides a useful tool for examining the persistence of candidate periodicities.

In this work, we applied the WWZ method to perform a time-frequency analysis of the candidate  {recurrent} signals with the period  {grid} range between 100 and 1500$\ {\rm d}$. The decay constant is set to $c = 1/8\pi^2 \approx 0.0125$ \citep{1996AJ....112.1709F}. To account for edge effects in the wavelet analysis, the regions significantly affected by window truncation were characterized by the cone of influence (COI) \citep{1998BAMS...79...61T}. When computing the time-integrated WWZ power spectrum, we excluded the power within the COI in order to reduce the influence of boundary effects on the period determination.

The resulting WWZ time-frequency map and the corresponding integrated power spectrum are shown in Figure \ref{fig:wwz}. For the C band, the WWZ power map shows a persistent enhancement around $\sim1000$~d over a substantial fraction of the observational baseline. The integrated spectrum excluding the COI shows a peak near $1014$~d and exceeds the $3\sigma$ local significance level. In the K band, the WWZ map also shows enhanced power around $\sim1000$~d, but the enhancement is weaker and less persistent. A substantial fraction of the response is affected by the COI, and the integrated spectrum excluding the COI does not show a statistically significant peak. Instead, it exhibits a broad plateau around the characteristic timescale. The behaviors in both bands are consistent with the GLSP response, indicating that the recurrent component is likely present in both bands. Again, we use this method primarily as a consistency check on the characteristic timescale rather than as an independent high-precision period measurement.

As a complementary time-domain test, we calculated the auto-correlation functions using the z-transformed discrete correlation function (ZDCF), introduced by \cite{1997ASSL..218..163A}. The first non-zero positive Auto-ZDCF peak occurs at $\tau_{\rm C}=993.02^{+22.40}_{-10.84}$~d in the C band and $\tau_{\rm K}=1059.59^{+34.53}_{-17.20}$~d in the K band. These quoted asymmetric ranges represent the widths of the corresponding ZDCF lag bins, rather than statistical uncertainties propagated from the measurement errors. The two characteristic lags are consistent with the $\sim1000$-d timescale obtained from the GLSP and WWZ analyses.

\begin{deluxetable*}{llcccc}
\tablewidth{0pt}
\tablecaption{Summary of the recurrent variability analysis of 4C+72.28 \label{tab:qpo_results}}
\renewcommand{\arraystretch}{1.3}
\tablehead{
\multirow{2}{*}{Method} & \multirow{2}{*}{Estimator} & \multicolumn{2}{c}{C-band (4.8 GHz)} & \multicolumn{2}{c}{K-band (23.6 GHz)} \\
\cmidrule(lr){3-4} \cmidrule(lr){5-6}
 & & Timescale [days] & Significance & Timescale [days] & Significance
 }
\startdata
\multirow{2}{*}{ GLSP}
 & Peak         & 1025.76                    & \multirow{2}{*}{\begin{tabular}{@{}c@{}} $3.29\sigma$ (local)\\ $1.95\sigma$, $1.82\sigma$ (global)\end{tabular}} & 1087.81                       & \multirow{2}{*}{\begin{tabular}{@{}c@{}} $2.69\sigma$ (local)\\ $2.18\sigma$, $1.77\sigma$ (global)\end{tabular}} \\
 & FR & $1027.97 \pm 3.31$         &                          & $1099.35 \pm 25.30$        &                          \\
\midrule
\addlinespace[3pt]
WWZ    &    & $1013.8 \pm 1.41$          & $>3\sigma$ (local)               & $-$                        & $-$                      \\
\midrule
Auto-ZDCF & & $993.02^{+22.40}_{-10.84}$ & $>2\sigma$ (local)               & $1059.59^{+34.53}_{-17.20}$ & $>1\sigma$ (local)       \\
\enddata
\tablecomments{The two GLSP rows share a single significance result; the local and global significances are evaluated on the strongest periodogram peak. \rev{The second global significance values for the GLSP are calculated over the same 100-1500\ d period range for both bands and are included only for cross-band comparison;} the band-specific data-driven intervals remain our primary global significance estimates as the first values shown. The GLSP FR estimates are obtained from $10^4$ flux randomization realizations and characterize the propagation of measurement uncertainties into the peak-period localization, which does not include the full method- and sampling-dependent uncertainty. Both the WWZ and Auto-ZDCF values represent the locations of the power/correlation peak, and the quoted ranges correspond to the frequency/lag-bin widths rather than measurement error uncertainties.}
\end{deluxetable*}


Taken together, the three analyses consistently identify a characteristic recurrent timescale close to $1000$~d, with the strongest and most persistent evidence found in the C band. The K-band data show weaker but broadly consistent responses in the GLSP, WWZ, and Auto-ZDCF analyses. The agreement among these methods is consistent with a recurrent variability component on this timescale and motivates describing it as a candidate quasi-periodic modulation. However, the Monte Carlo significance analysis does not establish a definitive detection. We therefore refer to the ~$1000$-d feature as a candidate recurrent modulation rather than a confirmed QPO throughout this work.

\section{Multi-timescale Variability Decomposition} \label{sec:mcmc}

The multiple analyses described above suggest a decomposition of the radio light curves into a long-term linear trend plus a short-timescale recurrent quasi-periodic modulation. These two behaviors motivate a simple phenomenological decomposition of each light curve into a linear trend plus a sinusoidal modulation. To quantify this decomposition and to test whether the recurrent component shares a common period across the two bands, we perform independent and joint Bayesian Markov Chain Monte Carlo (MCMC) fits to the dual-band light curves. The model form we use is
\begin{equation}
    F(t) = b + k\cdot t + A \sin(2\pi\cdot t/T + \phi),
\end{equation}
where $b$ is the flux normalization, $k$ represents the long-term evolution rate, $A$ is the modulation amplitude, $T$ is the characteristic timescale, and $\phi$ is the phase offset. We differentiate the two bands by assigning distinct subscripts to the parameters. For all the details in MCMC fitting, see Appendix~\ref{appendix: mcmc}.  
\rev{The MCMC fit is not an independent test for periodicity; under the linear-plus-sinusoidal decomposition it quantifies the parameter ranges consistent with the data. The resulting period uncertainties are conditional on this decomposition and do not constitute evidence for long-term phase coherence.}

\begin{figure}
    \centering
    \includegraphics[angle=0,scale=0.4]{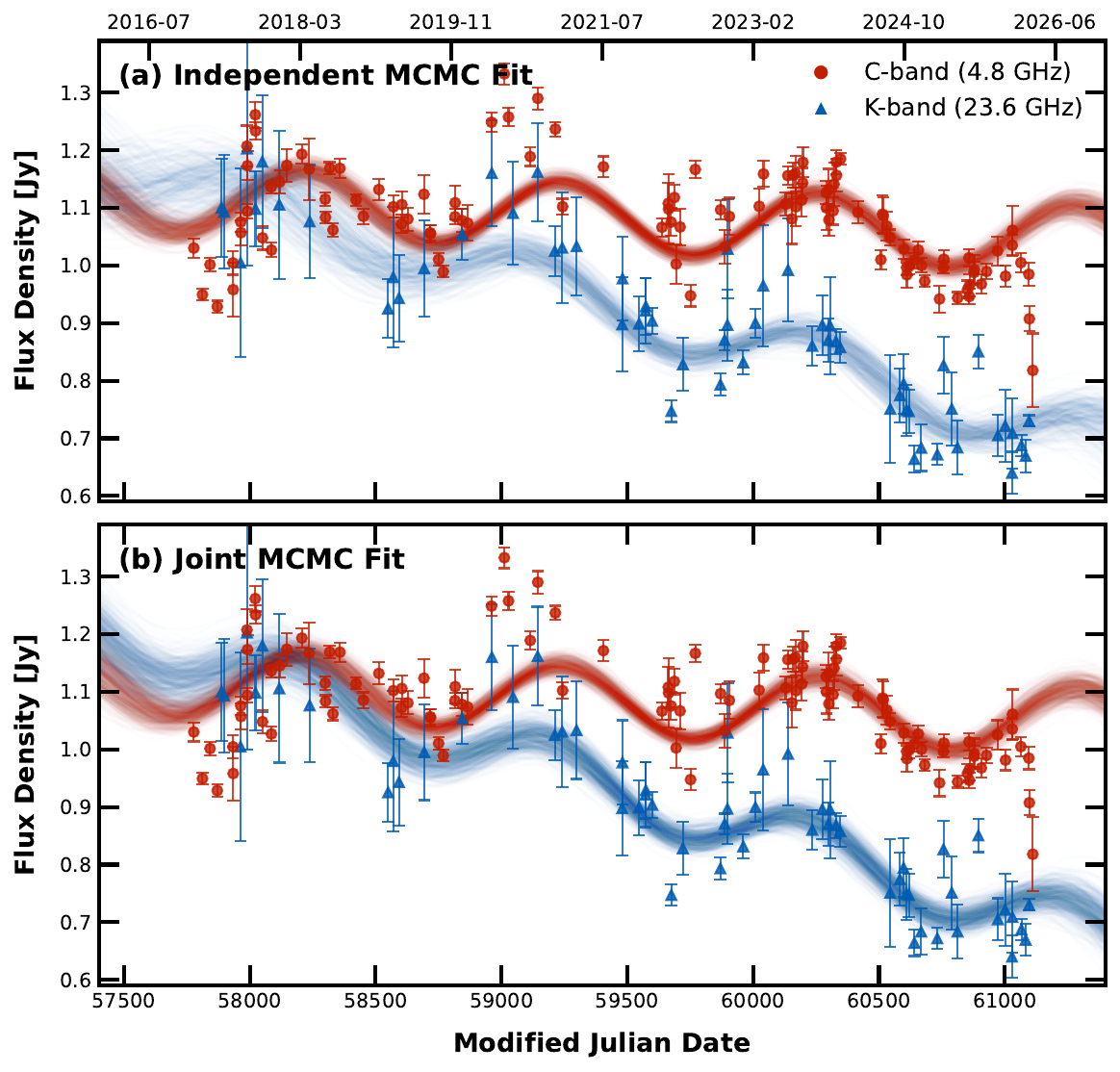}
    \caption{Posterior model realizations from the MCMC fits to the C- and K-band light curves. Panel (a) shows the results of independent fits, while panel (b) shows the results of the joint fit. The red circles and blue triangles represent the observed C- and K-band light curves, respectively. For each band and fitting case, 500 model curves are randomly selected from quasi independent samples of the converged MCMC posterior.}
    \label{fig:pm_mcmc}
\end{figure}

The independent fits listed in Table~\ref{tab:mcmc} return $T_{\rm C}=1020.8^{+18.2}_{-17.9}$\,d and $T_{\rm K}=1122^{+76}_{-77}$\,d, consistent with the GLSP, WWZ, and auto-ZDCF results in Table~\ref{tab:qpo_results}. The trend terms support the frequency-dependent secular evolution found in Section \ref{sec:lse}: the C-band slope corresponds to a fractional change of only $\approx6\%$ over the baseline, an order of magnitude weaker than the $\approx40\%$ decline of the K band. The fitted amplitudes, $A\approx0.06$\,Jy, correspond to $\approx6\%$ of the mean flux densities. The jitter terms provide an internal consistency check: $\sigma_{{\rm jit},{\rm C}}\approx0.07$\,Jy substantially exceeds the mean C-band measurement error ($\approx0.03$\,Jy), indicating that the C-band residuals are dominated by a stochastic process, in agreement with the anchored red noise background described in Appendix~\ref{appendix: rdn}, whereas $\sigma_{{\rm jit},{\rm K}}\approx0.05$\,Jy is comparable to the K-band measurement errors, consistent with its nearly white-noise-like PSD background.


The joint fit imposes a shared period across the two bands, which is a much stronger assumption of a common physical modulation on both bands. Under this assumption, it returns a common period of $T=1026.68$\,d, consistent with the independent C-band timing results. This should be interpreted as a consistency check under the shared-period hypothesis, rather than as independent evidence that a common physical modulation is present.
With a common modulation period, the joint posterior provides a stable inter-band phase offset,  {$\Delta\phi= -0.288^{+0.242}_{-0.253}$\,rad,} corresponding to $\Delta t=(\phi_{\rm C}-\phi_{\rm K})\,T/2\pi=-47.12_{-41.58}^{+39.57}$\,d (K leading), i.e., less than $5\%$ of the modulation time scale and with large uncertainty. The derived phase offset represents the relative timing of the recurrent component rather than a direct measurement of the light-curve time lag. Thus, we regard this as a small phase shift rather than a robust physical lag.

\begin{deluxetable*}{lcccc}
\tablecaption{Posterior parameters of the linear + sinusoidal MCMC fits to the C- and K-band light curves\label{tab:mcmc}}
\renewcommand{\arraystretch}{1.4}
\tablehead{
 & \multicolumn{2}{c}{Independent fit} & \multicolumn{2}{c}{Joint fit} \\
Parameter   & C band   & K band   & C band   & K band  
}
\startdata
$b$ [Jy]            & $1.0834^{+0.0054}_{-0.0053}$         & $0.9359^{+0.0104}_{-0.0104}$        & $1.0833^{+0.0053}_{-0.0053}$    & $0.9386^{+0.0103}_{-0.0104}$ \\
$k$ [$10^{-5}$Jy\,d$^{-1}$]  & $-1.9488^{+0.5099}_{-0.5160}$ & $-12.522^{+1.119}_{-1.197}$ & $-1.9087^{+0.5076}_{-0.5060}$ & $-13.363^{+1.029}_{-1.009}$ \\
$A$ [Jy]            & $0.0590^{+0.0074}_{-0.0073}$        & $0.0531^{+0.0114}_{-0.0118}$      & $0.0590^{+0.0073}_{-0.0072}$    & $0.0512^{+0.0113}_{-0.0111}$ \\
$T$ [d]             & $1020.80^{+18.17}_{-17.90}$            & $1122.00^{+75.98}_{-76.86}$                & \multicolumn{2}{c}{$1026.68^{+18.72}_{-17.71}$} \\
$\phi$ [rad]        & $2.557^{+0.1348}_{-0.1309}$           & $3.1256^{+0.2916}_{-0.2981}$         & $2.5654^{+0.1363}_{-0.1318}$    & $2.8558^{+0.2200}_{-0.2073}$ \\
$\sigma_{\rm jit}$ [Jy] &$0.0680^{+0.0042}_{-0.0038}$       & $0.0539^{+0.0072}_{-0.0068}$        & $0.0679^{+0.0041}_{-0.0038}$    & $0.0544^{+0.0072}_{-0.0066}$      \\
\enddata
\tablecomments{The model is $F(t)=b+kt+A\sin(2\pi t/T+\phi)$, with phases referenced to MJD\,59400. In the fits a jitter term $\sigma_{\rm jit}=e^{\ln{\rm jit}}$ is added in quadrature to the measurement uncertainties. Quoted values are posterior medians with 16/84 percentiles.}
\end{deluxetable*}

To evaluate whether the proposed decomposition is statistically acceptable, we use the information criterion to compare three phenomenological models: a linear model, a sinusoidal model, and a combined linear+sinusoidal model.

\begin{deluxetable}{llcccc}
\tablecaption{ {Information-criterion comparison of phenomenological models for the C- and K-band light curves}\label{tab:aicbic}}
\renewcommand{\arraystretch}{1.3}
\tablehead{Band & Model & $N_{\rm par}$ & $\chi^2_{\rm red}$ & AIC & BIC }
\startdata
  & Linear      & 2 & 1.349 & 275.09 & 281.72  \\
C & Sinusoid    & 4 & 1.086 & 224.04 & 237.30  \\
  & Linear+Sin  & 5 & 1.006 & 209.16 & 225.73  \\
\hline
  & Linear      & 2 & 1.043  & 140.59  & 146.38   \\
K & Sinusoid    & 4 & 3.165  & 416.29  & 427.86   \\
  & Linear+Sin  & 5 & 0.858  & 119.78  & 134.24   \\
\enddata
\tablecomments{The three models are $F(t)=b+kt$ (Linear), $F(t)=b+A\sin(2\pi t/T+\phi)$ (Sinusoid), and their superposition (Linear+Sin). $N_{\rm par}$ is the number of free parameters; the \(\chi^2\), AIC, and BIC values are computed using the effective uncertainty \(\sigma_{\rm eff}^{2}=\sigma_i^{2}+\sigma_{\rm jit}^{2}\). ${\rm AIC}=\chi^2+2N_{\rm par}$ and ${\rm BIC}=\chi^2+N_{\rm par}\ln N$, with $N=203$ (C) and $N=133$ (K). }
\end{deluxetable}


As shown in Table~\ref{tab:aicbic}, the \(\chi^2\), AIC, and BIC values are calculated using the same effective uncertainty adopted in the MCMC likelihood, \(\sigma_{\rm eff}^{2}=\sigma_i^{2}+\sigma_{\rm jit}^{2}\). We adopted fixed, band-specific jitter values equal to the posterior medians \(\sigma_{\mathrm{jit,C}}=0.068\) and \(\sigma_{\mathrm{jit,K}}=0.0539\) for all three candidate models within each band. The linear-plus-sinusoidal model yields the lowest information criteria in both bands, providing phenomenological support for decomposing the light curves into a secular trend and a recurrent component. The second-preferred model differs between bands: in the C band it is the pure sinusoidal model, whereas in the K band it is the linear model, consistent with the weaker recurrent signal in the K band. \rev{Because the residuals still contain unmodeled stochastic variability, these rankings are a relative comparison among three parameterized models and do not constitute independent evidence for a recurrent component; the primary evidence for the candidate timescale rests on the significance analysis in Section~\ref{sec:qpo}.}


\section{Cross-band Correlation Analysis} \label{sec:ccf}

After phenomenologically decomposing the variability in the 4C+72.28 radio light curves, the next key question is: Are the long-term spectral evolution and the recurrent variability produced by the same frequency-dependent process? If a single frequency-dependent process, such as a perturbation propagating in a stratified opaque jet, dominates both components simultaneously, then the cross-correlation between the original light curve and the detrended light curve should exhibit similar delay structures. On the contrary, if the recurrent modulation is approximately achromatic, such as geometrical modulation, then the original cross-correlation should present offsets in the long-term component, while the recurrent component should correlate near zero delay. Therefore, we carry out the comparison between the original cross-correlation and the detrended cross-correlation by ZDCF as a component-dependent cross-band correlation test.

We calculated the cross ZDCF for the original C-K pair and also for the linearly detrended C-K pair. Because the cross-correlation peak is broad and has no sharp peak, we did not use the single-point peak delay; instead, for each case, we generated 100,000 FR realizations and estimated the centroid delay of the central correlation hump (see Appendix~\ref{appendix:ccf}), the quoted asymmetric uncertainties correspond to the 16th and 84th percentiles of the resulting centroid lag distributions. For those realizations where the hump is too flat, we rejected them based on preset acceptance criteria. 

\begin{figure*}
    \centering
    \includegraphics[angle=0,scale=0.6]{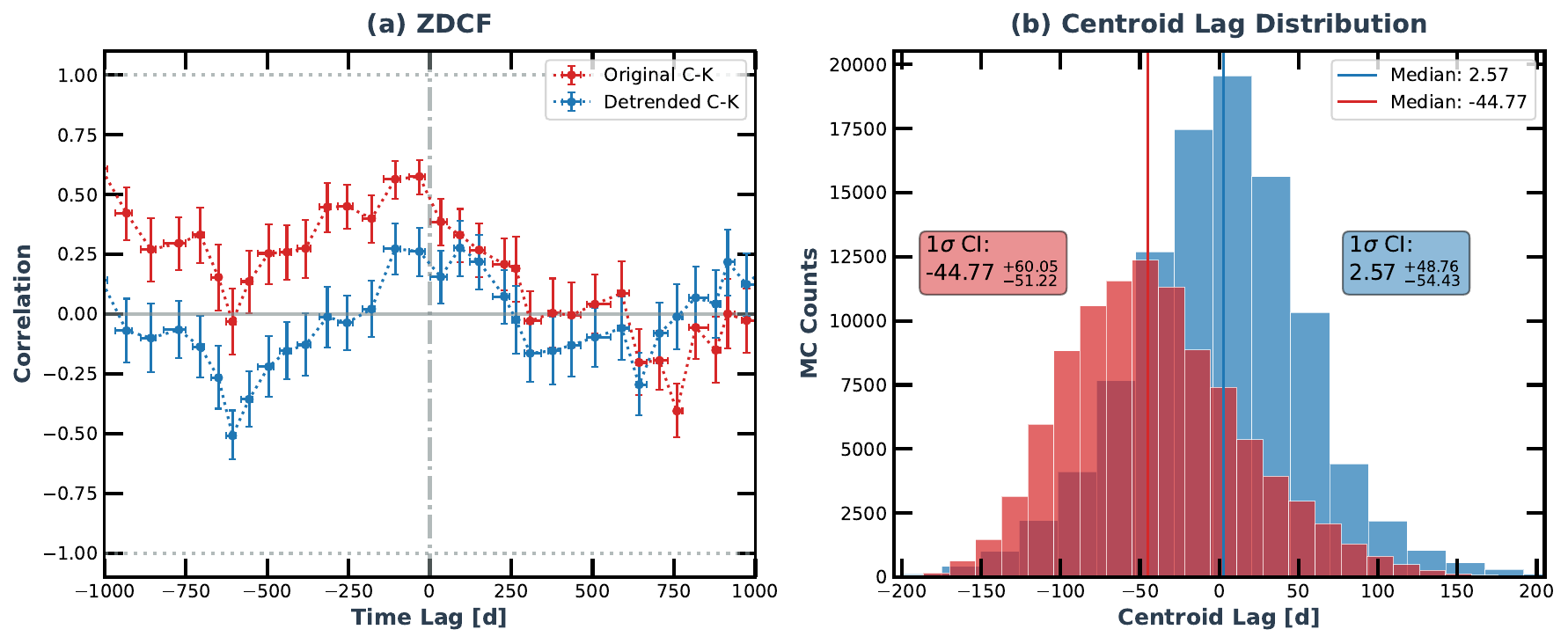}

    \caption{Cross correlation analysis results between the original and detrended C-K pairs. Subplot (a) shows the ZDCF of the original C-K light curves pair in red, and the detrended C-K pair in blue. Subplot (b) shows the 100,000 FR realizations distribution of the centroid lag of the central correlation peaks of the original and detrended C-K pairs.  {In both panels, negative lags indicate that the K-band leads the C-band variation.}}
    \label{fig:ccf}
\end{figure*}

The original C-K correlation shows a broad, offset structure with a centroid lag of $\tau_{\rm raw}=-44.77^{+60.05}_{-51.22}$\,d (negative lags indicate the K band leads). After removing the linear trends, the correlation power shows near zero lag, $\tau_{\rm det}=2.57^{+48.76}_{-54.43}$\,d. The two centroid distributions overlap substantially, we therefore cannot claim a significant detection of distinct lags. The qualitative result is that the correlation power shifts toward zero lag once the secular trend is removed, and the inter-band offset in the original light curves is likely dominated by the secular component, while the recurrent modulation behaves approximately simultaneously across the two bands (Figure ~\ref{fig:ccf}). \rev{The numerical similarity between the phase-derived offset and the centroid of the original C--K correlation does not make them independent measurements of the same physical lag, because the two quantities characterize different components of the variability.}

\section{Discussion} \label{sec:disc}

\subsection{Observational constraints on multi-timescale variability}

\rev{The radio monitoring of 4C+72.28 shows variability on more than one characteristic timescale. On the longer timescale, the C- and K-band light curves exhibit different secular trends, together with a gradual evolution of the C-K spectral index, indicating that the long-term variability is not purely achromatic.}

On shorter timescales, the detrended light curves show a \rev{candidate recurrent modulation} with a characteristic timescale of $\sim1000$ d. This feature is consistently identified by GLSP, WWZ, and ZDCF analyses, but the limited number of cycles (\(\approx3\)) and moderate significance ($3.29\sigma$ locally, $\sim 2 \sigma$ globally) imply that it should be regarded as a candidate quasi-periodic modulation rather than a firmly established periodicity.

Model comparison provides additional phenomenological support for this interpretation. In both the C and K bands, the linear+sinusoidal model yields the lowest reduced \(\chi^2\), AIC and BIC values, indicating that the inclusion of a recurrent component  {provides a better phenomenological description in both datasets.} However, the relative ranking of alternative models differs between bands: in the C band, the pure sinusoidal model is the second-best option, whereas in the K band the linear model ranks second. This difference suggests that the recurrent modulation is more clearly identifiable in the C band, while in the K band it is partially degenerate with the secular trend. 

Taken together, the spectral evolution, recurrent variability modulation, and model comparison \rev{are consistent with} a two-component description of the radio variability. This decomposition is empirical that it does not by itself imply physically independent emission regions, but rather provides a framework for interpreting distinct variability behaviors within a single system.

\subsection{Possible origin of the recurrent component}
The radio emission observed in the C and K-bands is expected to be dominated by synchrotron radiation from the relativistic jet \citep{1979ApJ...232...34B,2019ARA&A..57..467B}. The mechanisms capable of producing quasi-periodic radio variability can be broadly divided into geometrical and intrinsic processes. In a geometrical scenario, changes in the jet viewing angle modify the relativistic Doppler factor,
\begin{equation}
\delta = \left[\Gamma(1-\beta\cos\theta)\right]^{-1},
\end{equation}
where $\Gamma$ is the bulk Lorentz factor, $\beta c$ is the jet speed, and $\theta$ is the angle between the jet axis and the line of sight. For the convention $S_{\nu}\propto\nu^{\alpha}$, the observed synchrotron flux density scales approximately as
\begin{equation}
S_{\nu,\mathrm{obs}}\propto
\delta^{\,p-\alpha}S_{\nu,\mathrm{int}},
\end{equation}
where $p=2$ for a continuous jet and $p=3$ for a discrete moving component. Consequently, even a modest periodic change in viewing angle can produce a noticeable and approximately sinusoidal modulation of the observed flux density. Possible origins of such changes include jet precession, helical motion of emitting features \citep{1999A&A...347...30V}, binary-induced orbital or precessional motion, the Lense--Thirring effect \citep{1918PhyZ...19..156L,1998ApJ...492L..59S}, and magnetohydrodynamic instabilities such as kink modes.

The approximately sinusoidal appearance of the C-band recurrent component and the similar fractional amplitudes obtained from the joint MCMC fit provide a qualitative consistency check for such a scenario. At the reference epoch, the fitted fractional semi-amplitudes are
\begin{equation}
m_{\rm C}=\frac{A_{\rm C}}{b_{\rm C}}
=\frac{0.0590}{1.0833}\simeq0.0545,
\end{equation}
\begin{equation}
m_{\rm K}=\frac{A_{\rm K}}{b_{\rm K}}
=\frac{0.0512}{0.9386}\simeq0.0545 ,
\end{equation}
corresponding to approximately 5.45\% modulation in both the C and K bands. Thus, despite the substantially different secular slopes in the two bands, the relative amplitudes of the recurrent components are similar. For a sinusoidal modulation, the corresponding peak-to-trough flux-density ratio can be written as
\begin{equation}
R_{\nu}=\frac{1+m_{\nu}}{1-m_{\nu}},
\end{equation}
which gives $R_{\rm C}\simeq1.1152$ and $R_{\rm K}\simeq1.1154$. If the recurrent variability is produced predominantly by changes in relativistic beaming, with $S_{\nu,\rm obs}\propto\delta^{p-\alpha}$, the required Doppler-factor variation is
\begin{equation}
\frac{\delta_{\max}}{\delta_{\min}}
=R_{\nu}^{1/(p-\alpha)} .
\end{equation}

For representative radio spectral indices (-0.2$\lesssim\alpha\lesssim$0) and $p=2$--3, this corresponds to $\delta_{\max}/\delta_{\min}\simeq1.04$--1.06. Therefore, only a few-per-cent variation in the Doppler factor is sufficient to reproduce the observed recurrent amplitudes. This \rev{indicates} that geometrical modulation is physically plausible without requiring large changes in the jet orientation or bulk motion.

The similar fractional modulation in the two bands is thus qualitatively consistent with a common geometrical modulation, whereas their strongly different secular evolution likely requires an additional frequency-dependent contribution. This calculation is only an illustrative consistency test; without independent constraints on the bulk Lorentz factor and viewing angle, the present data do not uniquely determine the geometrical parameters of the jet.

The cross-band phase and correlation behavior is also qualitatively compatible with this interpretation. The joint sinusoidal model yields a small phase offset between the C- and K-band recurrent components, while the detrended ZDCF gives a centroid consistent with zero lag within the current uncertainties.  {The results are consistent with approximately simultaneous recurrent variability in the two bands, but do not independently establish a common physical modulation mechanism or long-term phase coherence.} Such behavior is naturally produced if both bands are affected by a common modulation of the observed Doppler factor. However, this interpretation is not unique. Intrinsic processes can also produce quasi-periodic variation through changes in the injected electron population, magnetic field, particle-acceleration efficiency, accretion rate, or shock formation and propagation \citep{1985ApJ...298..114M}.

The present data therefore do not distinguish uniquely between geometrical and intrinsic mechanisms for the recurrent component. In particular, the limited number of observed cycles and the modest statistical significance of the candidate timescale prevent us from treating the sinusoidal behavior as evidence for a specific periodic physical process. The geometrical interpretation should instead be regarded as a physically plausible explanation that is consistent with the observed waveform, fractional amplitude, and approximate cross-band phase coherence.

\subsection{Possible origin of the secular chromatic evolution}
In contrast to the recurrent component, the long-term variability exhibits a clear frequency dependence. The C-band flux density remains comparatively stable, whereas the K-band flux density shows a substantially stronger secular decline. The corresponding two-point C--K spectral index also shows a gradual long-term evolution (Figure~\ref{fig:lcs}), driven mainly by the decreasing contribution of the higher-frequency emission. This chromatic behavior indicates that the source variability cannot be described by a simple, purely achromatic modulation of a single stationary spectral component. The K-band emission is expected to arise from a more compact and generally more upstream region of the jet than the C-band emission. Its stronger decline may therefore reflect fading of a compact high-frequency synchrotron component, changes in particle injection or magnetic field, radiative and adiabatic evolution of the emitting plasma, or changes in the opacity structure of the jet. Evolution of the synchrotron self-absorption or external opacity structure may also alter the relative contributions of the C- and K-band emitting regions. 

An alternative possibility is that the secular evolution and recurrent modulation are both produced by the same underlying jet process but affect different portions of the stratified radio-emitting region. In this case, frequency-dependent opacity, spatially varying particle populations, or different characteristic cooling times could produce different long-term responses even if the short-timescale modulation is common to the two bands. The present observations cannot distinguish this scenario from genuinely distinct physical drivers.

\subsection{A possible two-component physical picture}
The combined observational results \rev{are consistent with} a picture in which the radio variability of 4C+72.28 contains a slow chromatic component and a recurrent component with a characteristic timescale of approximately $1000$ d. The two components exhibit different observational properties: the secular component is associated with the long-term C--K spectral evolution, whereas the recurrent component is more clearly identified through the quasi-periodic analyses and is approximately simultaneous in the two radio bands under the joint MCMC model. The preference for the linear+sinusoidal model in both bands provides an  {additional} phenomenological consistency check for this decomposition.

The ZDCF comparison provides a complementary constraint on this picture. The original C--K light curves show a broad correlation structure with a nominal K-leading centroid lag of $\tau_{\rm raw}=-44.77^{+60.05}_{-51.22}$~d, whereas after removing the secular trends the correlation is centered close to zero lag, with $\tau_{\rm det}=2.57^{+48.76}_{-54.43}$~d (Figure~\ref{fig:ccf}). The corresponding lag distributions overlap substantially, and therefore the difference between the two cases is not statistically significant. Nevertheless, the shift of the correlation centroid toward zero lag after detrending is qualitatively consistent with the recurrent components varying approximately simultaneously in the two bands, while the offset present in the original light curves is likely influenced by the frequency-dependent secular evolution. 

\rev{This phenomenological decomposition is not evidence for two physically independent mechanisms or emission regions. The observations show that the long-term and recurrent components have different variability properties and different frequency dependences.} A plausible interpretation is that the recurrent component is associated with a coherent modulation of the relativistic jet, potentially through Doppler-factor variations, while the slower component reflects intrinsic evolution and/or opacity changes in the radio-emitting plasma. Under this interpretation, the observed radio light curve represents the superposition of a common recurrent modulation and a slower evolution of the underlying synchrotron spectrum.

\subsection{Implications and future tests}
The observed timescale is comparable to year-scale radio QPO candidates reported in several other blazars, including AO\,0235+164, OT\,081, PKS\,2131$-$021, S5\,0716+714, and PKS\,J2156$-$0037 \citep{2021MNRAS.501.5997T,2021MNRAS.506.1540L,2022ApJ...926L..35O,2023ApJ...943..157L,2024MNRAS.531.3927M}. Some of the most compelling radio cases are supported by substantially longer observing baselines, multiple radio frequencies, or repeated intervals of phase-coherent modulation. For example, the long-term radio record of PKS\,2131$-$021 contains two separated episodes with mutually consistent periods and phases \citep{2022ApJ...926L..35O}. In comparison, the present candidate is based on approximately three cycles and is statistically strongest only in the C band. This comparison reinforces the need for continued monitoring before the persistence and phase coherence of the $\sim1000\ {\rm d}$ modulation can be established.

A possible but currently unconstrained scenario is a binary-supermassive-black-hole origin, which could in principle produce periodic modulation through orbital motion, periodic accretion, or binary-induced jet precession. However, the redshift of 4C+72.28 remains unknown, preventing conversion of the observed period into a source-frame timescale and limiting estimates of its luminosity, physical jet scale, and any binary orbital parameters. The absence of a reliable black-hole mass estimate and of direct evidence for periodic structural changes also prevents meaningful constraints on a binary model. The current radio light curves alone therefore do not provide sufficient evidence to favor a binary interpretation over other geometrical or intrinsic jet processes.

Continued observations will be essential for distinguishing among these scenarios. If the modulation is predominantly geometrical, future cycles should preserve an approximately stable phase, and multi-frequency light curves may show broadly simultaneous variations. Multi-epoch VLBI observations could reveal recurrent changes in the jet position angle, core structure, or trajectories of individual components, while polarimetric monitoring may detect associated changes in polarization degree or electric-vector position angle. Conversely, recurrent intrinsic disturbances should be accompanied by frequency-dependent time delays, stronger spectral evolution, changes in polarization, or the emergence of new VLBI components. Additional radio and millimeter-band monitoring, together with optical, infrared, and $\gamma$-ray light curves where sufficiently sampled, will be important for determining whether the candidate timescale persists and whether it is shared by other emission regions.

Such observations will be essential for determining whether the phenomenological decomposition identified here corresponds to genuinely distinct physical processes or instead reflects different timescales of variability within a single evolving relativistic jet.

\section{Summary}

We have used approximately nine years of 4.8 and 23.6\,GHz monitoring from the SMMAN project to investigate the multi-timescale radio variability of the blazar 4C+72.28. Our main results are as follows:

\begin{itemize}

\item The radio variability exhibits a slow, frequency-dependent secular component, manifested by the substantially stronger long-term decline at K band and the corresponding evolution of the C--K spectral index, together with a  {candidate} quasi-periodic modulation on a characteristic timescale of approximately $1000$\,d.

\item The recurrent timescale is consistently identified in the C-band timing analyses, although its global statistical significance remains modest. Independent MCMC fits recover consistent timescales in both bands, while the joint fit gives $T=1026.68^{+18.72}_{-17.71}$\,d and similar fractional recurrent amplitudes of $\simeq5.45\%$. Information-criterion comparisons favor the linear-plus-sinusoidal model in both bands, supporting the phenomenological decomposition into secular and recurrent variability.

\item The C--K ZDCF yields a broad centroid lag of $\tau_{\rm raw}=-44.77^{+60.05}_{-51.22}$\,d for the original light curves and $\tau_{\rm det}=2.57^{+48.76}_{-54.43}$\,d after detrending. Although the difference is not statistically significant, the shift toward zero lag is qualitatively consistent with a recurrent component that is more nearly coherent across the two bands than the total long-term variability.

\item The combined results  {are consistent with} a picture in which a quasi-periodic modulation is superposed on a slower chromatic evolution of the radio-emitting jet.  {Geometrical Doppler-factor modulation is one plausible origin of the recurrent component, although correlated intrinsic or opacity-related variability cannot be excluded, whereas the secular component most likely reflects intrinsic plasma evolution and/or opacity changes in the radio-emitting plasma.} The present data do not uniquely distinguish these scenarios, and continued multi-frequency monitoring, together with VLBI and polarimetric observations, is required to establish the persistence and physical origin of the recurrent variability.

\end{itemize}

\begin{acknowledgments}
We thank the anonymous reviewer for constructive comments, which imporved the quality and clarity of the manuscript.
We thank Alok C. Gupta, Ashutosh Tripathi, Ming Zhang, and Xi Yan for valuable discussions on the data analysis methods. This work has made use of the SMMAN data observed by the Nanshan 26-m Radio Telescope (NSRT), operated by the Xinjiang Astronomical Observatory (XAO), CAS. This work is supported by the Tianshan Talent Training Program (Grant No. 2023TSYCCX0099) and the National Key R\&D Program of China (Grant No. 2024YFA1611503). Z.H. is supported by the Xinjiang Science and Technology Program (Grant No. 2024D03008). This work is also partly supported by the Urumqi Nanshan Astronomy and Deep Space Exploration Observation and Research Station of Xinjiang (XJYWZ2303), the Central Guidance for Local Science and Technology Development Fund (Grant No. ZYYD2026JD01), and the Astrometric Reference Frame project (Grant No. JZZX-0102). 
\end{acknowledgments}

\facilities{NSRT (4.8 GHz and 23.6 GHz)}
\software{\texttt{DELightcurveSimulation} \citep{2016ascl.soft02012C}, \texttt{emcee} \citep{2013PASP..125..306F}}

\appendix

\section{Stochastic background modeling} \label{appendix: rdn}
For real monitoring light curves, the background power spectrum theoretically consists of a red-noise component corresponding to a source-intrinsic stochastic process and a white-noise component arising from the measurement fluctuation. The simplest description of red noise is a power-law function, whose slope, i.e., the spectral index, reflects whether the variability is dominated by long-term trends or by short-timescale fluctuations. White noise, on the other hand, exhibits constant power across all frequencies and can be directly described by a constant term in the power spectrum.

Modeling the red-noise background of AGN light curves is well established. For example, \cite{2005A&A...431..391V} systematically discussed red-noise light curves and the statistical tests for periodicity. The REDFIT method models the red-noise background using a first-order autoregressive process, namely an AR(1) process \citep{2002CG.....28..421S}. More general CARMA(p,q) models can describe more complex continuous-time stochastic processes \citep{2014ApJ...788...33K}. However, in practical data analysis, if the white-noise level is not properly constrained, the parameters of the red-noise model may be affected by high-frequency measurement noise, especially when the data are unevenly sampled, the measurement uncertainties are large, or the monitoring baseline is limited. Therefore, we first independently estimated the white-noise level caused by measurement uncertainties, and then constrained it in the subsequent PSD fitting. This treatment reduces the degeneracy and gives clear physical interpretation of the red-noise background. 

To estimate the white-noise level caused by the measurement uncertainties, we generated 10,000 pure measurement-error realizations for each band. \rev{Each realization keeps the actual observing epochs fixed;} at each epoch, a flux density value is drawn from a Gaussian distribution whose mean is a constant equal to the time-averaged flux density of the light curve and whose standard deviation is the measurement uncertainty of that epoch. Since every epoch shares the same constant mean, these realizations contain only the heteroscedastic measurement noise and no intrinsic source variability, and their power spectra are therefore expected to be flat. We calculated the unnormalized PSD of each realization, fitted it with a constant, and took the mean and standard deviation of the 10,000 best-fitting constants as the white-noise level and its uncertainty. The resulting white-noise power level of the unnormalized PSD is $(1.14\pm0.18)\times 10^{-2}\ {\rm Jy^2\,d}$ for the C-band light curve, and $(2.45\pm0.55)\times 10^{-1}\ {\rm Jy^2\,d}$ for the K-band light curve.

After obtaining the white-noise levels of the C-band and K-band light curves, we then modeled the PSDs of the target light curves. Considering the data quality and model complexity, we described the spectrum using a broken power-law for the red-noise component and a constant white-noise term. The model can be written as
\begin{equation}
\mathbf{P}(f) = 
\begin{cases} 
\mathbf{P}_{\text{white}} \left( {f}/{f_b} \right)^{-\alpha}, & \text{for } f \leq f_b, \\
\mathbf{P}_{\text{white}}, & \text{for } f > f_b.
\end{cases}
\end{equation}
where $f_b$ is the breaking frequency of the broken power-law, $\alpha$ is the red noise power-law spectral index, and $ \mathbf{P}_{\rm white}$ represents the white-noise power. In the actual fitting, $\mathbf{P}_{\rm white}$ is constrained within $2\sigma$ range of the estimate results above. Compared with a fully free broken power-law model, this model provides better convergence and allows a more stable characterization of the intrinsic red-noise background of the target light curves.

The constrained broken power-law model provides stable fits to the unnormalized PSDs of both bands. For the C band, the best-fitting parameters are $\alpha = 0.89$, $f_b = 0.070\ {\rm d}^{-1}$, and $\mathbf{P}_{\rm white} = 0.011\ {\rm Jy^2\,d}$, consistent with the independently estimated white-noise level and therefore supporting the anchoring strategy. For the linearly detrended K band, the best fit gives  $\alpha = 0.05$, $f_b = 0.011\ {\rm d}^{-1}$, and $\mathbf{P}_{\rm white} = 0.180\ {\rm Jy^2\,d}$. The nearly zero slope indicates that the K-band PSD is close to pure white noise over the searched frequency range, consistent with the fact that the LS-fit sinusoidal amplitude (about 0.066 Jy) is comparable to the mean K-band measurement uncertainty (about 0.079 Jy). This explains why the K-band null hypothesis is nearly white and why its periodic signal is more difficult to detect. While the C-band fitted sinusoidal amplitude and the mean measurement uncertainty is about 0.079 Jy and 0.027 Jy, respectively. 

\rev{Based on the fitted background models above, we constructed the stochastic processes that provide the null hypothesis for the significance analysis.} The PSD fitting method used above follows the workflow of the \texttt{DELightcurveSimulation}\footnote{\url{https://github.com/samconnolly/DELightcurveSimulation}} package \citep{2016ascl.soft02012C}.

\section{Significance Analysis} \label{appendix: sig}
To robustly evaluate the statistical significance of our analysis results, in this work we follow the Monte Carlo simulation-based method of \cite{2013MNRAS.433..907E} (hereafter E13), and use the Python package \texttt{DELightcurveSimulation}, coded by \cite{2016ascl.soft02012C}, to generate simulated light curves. The E13 method performs Monte Carlo simulations by matching both the power spectral density (PSD) and probability density function (PDF) of the original target light curve and generates stochastic process light curves with the best-fitting red noise background model estimated in the previous subsection. And then we repeat exactly the same data-analysis procedures as those applied to the target light curve, to obtain the empirical distributions of the relevant statistics under the null hypothesis. Here, the null hypothesis is that the observed light curve is produced only by a stationary stochastic process with pure red + white noise and contains no additional quasi-periodic components. 

In this work, based on the best-fitting background PSD model obtained in the previous subsection, we generated 100,000 simulated random light curves. We used exactly the same GLSP frequency grid, WWZ frequency grid, and minimum number of data points per ZDCF bin as adopted for the observed light curves, thereby constructing the distributions of the corresponding statistics under the stochastic process null hypothesis. At each frequency or time-lag grid point, the corresponding percentiles were calculated from the distributions of the 100,000 simulations, from which the frequency- or time-lag-dependent local significance curves were constructed. 

For the GLSP and the integrated WWZ power spectrum after excluding the COI, we calculated the single-tailed 84.135\%, 97.725\%, and 99.865\% percentile levels, corresponding to the local significance thresholds of approximately $1\sigma$, $2\sigma$, and $3\sigma$, respectively. For the auto-ZDCF, since both positive and negative deviations are statistically meaningful, we calculated the double-tailed percentile pairs of $(15.865\%,\,84.135\%)$, $(2.275\%,\,97.725\%)$, and $(0.135\%,\,99.865\%)$ as the corresponding envelopes. These thresholds can be used to evaluate the local significance of a candidate peak at a specified location. 

The above method provides the local significance, which shows the frequency-dependent effects of limited baselines, uneven sampling, red noise leakage and spectral windows. It evaluates the probability that a stochastic process under the null hypothesis produces a statistical fluctuation no weaker than the observed signal at a specified frequency or time lag. Since a practical periodicity search usually searches for the most significant peak over a frequency interval, the look-elsewhere effect \citep{2010EPJC...70..525G} also needs to be considered. In other words, even if the probability for a red-noise process to produce a high-power peak at one specified frequency is low, it may still produce an equally or more significant random peak at another frequency within the whole search interval. The global significance is used to quantify the probability that at least one equally or more significant peak appears anywhere within the entire pre-defined search interval, and is therefore generally more conservative than the local significance. 

The global significance highly depends on the pre-defined frequency search interval. The complete frequency grid used for displaying the GLSP covers a relatively broad frequency range, but its edge regions may not be suitable for a reliable periodicity test. At the low-frequency end, the corresponding periods may last less than 2 cycles within the observational baseline, making them difficult to distinguish from long-term trends. At the high-frequency end, the results may be significantly affected by uneven sampling. 

Therefore, according to the observational duration and the typical sampling interval, by a data-driven rule (2 cycles to 10 cadences), we empirically adopt an effective frequency range for evaluating the global significance between $f_{\rm min} = {2}/{T}$ and $f_{\rm max} = {1}/{10\bar{\tau}}$,
where $T$ is the total observational duration of the light curve and $\bar{\tau}$ is the mean sampling interval. The corresponding effective period-search range is $10\bar{\tau} < P < {T}/{2}$. 

For the C band, this range is approximately $165~{\rm d} \leq P \leq 1667~{\rm d}$, while for the K band it is approximately $243~{\rm d} \leq P \leq 1605~{\rm d}$. This choice requires that the longest candidate period covers at least approximately two complete cycles within the observational baseline, while the shortest candidate period contains approximately ten typical sampling intervals. This is an empirical and conservative criterion designed to reduce the effects of the finite observational baseline and sparse, irregular sampling, rather than a universal strict boundary.

For the detailed calculation, we fully follow the method of \cite{2022ApJ...926L..35O}. First, at the frequency of the observed peak, we counted the number of simulated powers higher than the observed peak power, denoted by $N(\bf{P}_{\rm peak, sim} > \bf{P}_{\rm peak, obs})$. The single-period $p$-value of the observed peak was then calculated as 
\begin{equation}
    p_{\rm loc} = \frac{1 + N(\bf{P}_{\rm sim} > \bf{P}_{\rm obs})}{N_{\rm sim}+1}.
\end{equation}
Equivalently, the corresponding local significance is $\bf{S}_{\rm loc} = 1 - p_{\rm loc}.$

We then searched for the strongest GLSP peak within the pre-defined period-search range for each of the 100,000 simulated light curves, and calculated the local significance of every simulated GLSP peak. We further counted the number of simulations whose single-period $p$-value was smaller than that of the observed GLSP peak, denoted by $N(p_{\rm loc,sim}<p_{\rm loc,obs})$. The global FAP, or global $p$-value, was then calculated as 
\begin{equation}
    p_{\rm glb} = \frac{1+N(p_{\rm loc,sim}<p_{\rm loc,obs})}{N_{\rm sim}+1},
\end{equation}
where $N_{\rm sim}=100,000$. The global significance was therefore defined as $\bf{S}_{\rm glb} = 1 - p_{\rm glb}$.

\section{ {Robustness tests: Detrending and stochastic background modeling}}
\label{appendix:robustness}

The significance estimates in Section~\ref{sec:qpo} depend on both the preprocessing of the light curves and the stochastic background adopted for the Monte Carlo simulations. To test the robustness of the candidate recurrent modulation, we performed two additional experiments. The first examines the effect of linear detrending on the GLSP candidate. The second examines the sensitivity of the candidate false-alarm probabilities to alternative stochastic-background models.

\subsection{Effect of Detrending}

To examine whether the candidate is produced by the asymmetric detrending adopted in the main analysis, we repeated the same GLSP and significance analysis for the raw and linearly detrended C- and K-band light curves. The significance definitions are identical to those described in Section 4 and Appendix B. The results are listed in Table~\ref{tab:detrend_comparison}. \rev{The significance values in this table are based on $10^4$ MC realizations, whereas the fiducial values in the main text and in Table~\ref{tab:qpo_results} are based on $10^5$ realizations. The global significances quoted here therefore differ from Table~\ref{tab:qpo_results} by up to $\sim0.05\sigma$, which is consistent with the MC standard error of a $10^4$-realization estimate. These $10^4$-realization robustness runs are intended for computational efficiency and do not supersede the fiducial values.}

For the C band, the candidate peak remains close to 1000 d regardless of whether the weak linear trend is removed. The peak period changes from 1025.76 d in the raw light curve to 1010.22 d after detrending, while the local significance decreases from $3.29\sigma$ to $2.97\sigma$ and the global significance decreases from $1.97\sigma$ to $1.60\sigma$. Still, the C-band candidate is reasonably stable with respect to the tested detrending procedure.

For the K band, the raw light curve is dominated by a strong secular decline. Without detrending, this long-term trend increases the total variance of the light curve and therefore reduces the normalized GLSP peak power: the raw K-band peak power is 0.191, whereas the detrended K-band peak power is 0.262. The peak period also shifts, from 1188.54 d in the raw data to 1087.81 d after detrending. The raw K-band light curve yields a higher local significance ($3.01\sigma$) than the detrended light curve ($2.69\sigma$), but it does not yield a higher global significance. This behavior reflects the fact that detrending changes both the variance normalization and the fitted stochastic background, and hence the frequency-dependent local significance threshold.

More importantly, the Monte Carlo framework of \cite{2013MNRAS.433..907E} is based on the assumption of the light curve being approximately stationary. The raw K-band light curve violates this assumption because of its strong secular trend. We therefore do not regard the raw K-band significance as a reliable null-hypothesis result. The detrended K-band light curve is used in the main analysis because this choice satisfies the stationarity requirement more closely. This asymmetric preprocessing is therefore motivated by the different trend strengths in the two bands, rather than by the resulting significance values.


\renewcommand{\arraystretch}{1.3}
\begin{deluxetable*}{lccccc}
\tablecaption{GLSP peak statistics for the raw and linearly detrended C- and K-band light curves.\label{tab:detrend_comparison}}
\tablehead{\colhead{Light Curve}        &\colhead{Peak Period (d)}      &\colhead{Peak Power}      &
           \colhead{$\bf{S}_{\rm loc}$} &\colhead{$\bf{S}_{\rm glb}$}   &\colhead{$\bf{S}_{\rm glb}^{*}$}}
\startdata
Raw C band       &$1025.76$ &$0.428$ &$3.29\sigma$ ($0.05\%$) &$1.97\sigma$ ($2.45\%$) &$1.82\sigma$ ($3.47\%$) \\
Detrended C band &$1010.22$ &$0.369$ &$2.97\sigma$ ($0.15\%$) &$1.60\sigma$ ($5.44\%$) &$1.46\sigma$ ($7.22\%$) \\
Raw K band       &$1188.54$ &$0.191$ &$3.01\sigma$ ($0.13\%$) &$2.03\sigma$ ($2.13\%$) &$1.68\sigma$ ($4.63\%$) \\
Detrended K band &$1087.81$ &$0.262$ &$2.69\sigma$ ($0.36\%$) &$2.13\sigma$ ($1.65\%$) &$1.77\sigma$ ($3.82\%$) \\
\enddata

\tablecomments{Peak power is the normalized GLSP power.
Each entry gives the Gaussian-equivalent significance, with the corresponding one-sided false-alarm probability in parentheses.
$\bf{S}_{\rm loc}$ is the local significance at the listed peak frequency.
$\bf{S}_{\rm glb}$ is the global significance evaluated over the band-specific data-driven search range described in Appendix~\ref{appendix: sig}.
$\bf{S}_{\rm glb}^{*}$ is evaluated over the common period range $100$--$1500$\,d for both bands and is intended only for cross-band comparison.
Detrending removes the least-squares linear trend from each light curve.
The raw K-band result is shown for completeness; the detrended K-band light curve is used in the main analysis because the raw K-band data are dominated by the strong secular decline.
\rev{The significance values in this table are based on $10^4$ MC realizations, therefore, they differ from the $10^5$-realization fiducial results reported in Table~\ref{tab:qpo_results}.}
}
\end{deluxetable*}
\onecolumngrid

\subsection{Sensitivity to the Stochastic Background Model}

\renewcommand{\arraystretch}{1.1}
\begin{deluxetable*}{llcccc}
\tablecaption{ {Maximum-likelihood fits of stochastic background models to the raw C-band and detrended K-band PSDs.}
\label{tab:psd_background_models}}

\tablehead{\colhead{Model}              &\colhead{Parameter}          &\colhead{Best-fit}          &
           \colhead{Median${}^{+}_{-}$} &\colhead{Residual$^{\rm a}$} &\colhead{FAP$^{\rm b}$ (\%)}}

\startdata
\cutinhead{\it Raw C band}
Power law    & $A$              &$1.150\,\times10^{-3}$ &$1.007^{+0.528}_{-0.374}\,\times10^{-3}$ &$1.103\,\times10^{-1}$ &$0.02/1.05$ \\
             & $\alpha$         &$0.879$                &$0.866^{+0.104}_{-0.096}$                &                       &            \\
Bending PL   & $A$              &$1.520\,\times10^{-4}$ &$4.672^{+5.078}_{-2.912}\,\times10^{-4}$ &$1.018\,\times10^{-1}$ &$0.09/3.18$ \\
             & $\alpha$         &$1.182$                &$0.999^{+0.149}_{-0.135}$                &                       &            \\
             & $P_{\rm white}$  &$2.009\,\times10^{-2}$ &$0.840^{+1.002}_{-0.626}\,\times10^{-2}$ &                       &            \\
Broken PL    & $f_{\rm b}$      &$6.136\,\times10^{-2}$ &$6.775^{+2.343}_{-1.505}\,\times10^{-2}$ &$1.095\,\times10^{-1}$ &$0.04/1.98$ \\
             & $\alpha$         &$0.879$                &$0.870^{+0.100}_{-0.095}$                &                       &            \\
             & $P_{\rm white}$  &$1.294\,\times10^{-2}$ &$1.148^{+0.172}_{-0.172}\,\times10^{-2}$ &                       &            \\
Broken PL (Connolly)& $f_{\rm b}$      &$7.0\,\times10^{-2}$ & -                                  &$1.088\,\times10^{-1}$ &$0.05/2.56$ \\
             & $\alpha$         &$0.890$                & -                                       &                       &            \\
             & $P_{\rm white}$  &$1.10\,\times10^{-2}$  & -                                       &                       &            \\
\cutinhead{\it Detrended K band}
Power law    & $A$              &$4.037\,\times10^{-2}$ &$3.657^{+2.975}_{-1.847}\,\times10^{-2}$ &$1.537\,\times10^{-1}$ &$1.38/5.55$ \\
             & $\alpha$         &$0.273$                &$0.254^{+0.145}_{-0.122}$                &                       &            \\
Bending PL   & $A$              &$4.329\,\times10^{-2}$ &$1.590^{+3.194}_{-1.309}\,\times10^{-2}$ &$1.524\,\times10^{-1}$ &$1.36/5.42$ \\
             & $\alpha$         &$0.253$                &$0.244^{+0.232}_{-0.582}$                &                       &            \\
             & $P_{\rm white}$  &$0.100\,\times10^{-2}$ &$6.775^{+7.408}_{-6.272}\,\times10^{-2}$ &                       &            \\
Broken PL    & $f_{\rm b}$      &$1.520\,\times10^{-2}$ &$0.236^{+0.934}_{-0.120}\,\times10^{-2}$ &$1.584\,\times10^{-1}$ &$0.60/2.76$ \\
             & $\alpha$         &$0.152$                &$0.341^{+0.701}_{-0.290}$                &                       &            \\
             & $P_{\rm white}$  &$1.500\,\times10^{-1}$ &$1.598^{+0.157}_{-0.088}\,\times10^{-1}$ &                       &            \\
Broken PL (Connolly)& $f_{\rm b}$      &$1.1\,\times10^{-2}$ & -                                  & $1.670\,\times10^{-1}$&$0.36/1.46$ \\
             & $\alpha$         &$0.050$                & -                                       &                       &            \\
             & $P_{\rm white}$  &$1.80\,\times10^{-1}$  & -                                       &                       &            \\
\enddata
\tablenotetext{a}{Mean absolute residual between the observed PSD and the best-fit model, computed in linear PSD space.}
\tablenotetext{b}{Local/global false-alarm probabilities of the candidate GLSP peak evaluated under the corresponding fitted background model. The fiducial Broken PL (Connolly) model uses $10^5$ Monte Carlo realizations as adopted in the main text, while the alternative models use $10^4$ realizations for computational efficiency in this sensitivity test.}
\tablecomments{Best-fit values for the first three models are the maximum-likelihood values on the adopted parameter grid. Median values are the 50th percentiles; the superscripted uncertainties correspond to the 84th$-$50th and 50th$-$16th percentile intervals. $A$ and $P_{\rm white}$ are in units of Jy$^{2}$\,d, $f_{\rm b}$ is in d$^{-1}$, and $\alpha$ is dimensionless. The broken power law consists of a low-frequency power-law component with index $\alpha$ that flattens to the high-frequency level $P_{\rm white}$. The bending power law contains a power-law component plus an additive white-noise term $P_{\rm white}$. Global FAPs are computed using the same band-specific search ranges as in the main GLSP analysis. Because periodogram bins are correlated, the residual should be interpreted as a descriptive goodness-of-fit diagnostic rather than a formal $\chi^{2}$. The `Broken PL (Connolly)' row reports the parameters and significance results obtained using the \texttt{DELightcurveSimulation} package, which serves as the fiducial model in the main text. Since the fitting procedure does not provide marginalized posterior distributions, the Median column is left blank for this model.}
\end{deluxetable*}
\onecolumngrid

The fiducial stochastic background used in the main analysis is the white-noise-anchored broken power-law model described in Appendix~\ref{appendix: rdn}. This model was chosen primarily because the independent white-noise constraint stabilizes the separation between correlated red noise and high-frequency measurement noise for the two bands, which have substantially different measurement-noise levels. It was not chosen because it maximizes the candidate significance.

To test the sensitivity of the candidate significance to this modeling choice, we fit three background models to the raw C-band and detrended K-band PSDs: a single power law, $P(f)=A f^{-\alpha}$; the anchored broken power-law model of Appendix~\ref{appendix: rdn}; and a bending power-law model with a free additive white-noise term, $P(f)=A f^{-\alpha}+P_{\rm white}$. For each model, we obtained maximum-likelihood parameter estimates and used the 16th and 84th percentiles of the marginalized likelihood distributions as approximate $1\sigma$ uncertainties. The results are given in Table~\ref{tab:psd_background_models}.

Note that the maximum-likelihood fits performed in this section are independent of the background fitting used in the main analysis, which follows the \texttt{DELightcurveSimulation} workflow. The purpose of this independent fitting is to assess the sensitivity of the candidate significance to the choice of stochastic background model, rather than to replace the fiducial background. Accordingly, the $10^{4}$ Monte Carlo realizations used here are intended only for this model-comparison sensitivity test; the fiducial significance estimates reported in the main text remain based on $10^{5}$ realizations generated from the Appendix~\ref{appendix: rdn} background model.

Because individual periodogram ordinates do not have independent Gaussian errors, we do not assign formal $\chi^2$ goodness-of-fit statistics. Instead, we use the mean absolute residual between the observed PSD and the best-fit model, computed in linear PSD space, as a descriptive goodness-of-fit diagnostic. Table~\ref{tab:psd_background_models} shows that the residuals of the three models are broadly comparable in the C band, while the anchored broken power-law model provides a stricter high-frequency constraint in the K band which leads to a slightly larger residual than other two, less constrained models.

The comparison also shows that the breaking frequency and the white-noise amplitude are only weakly constrained when the white-noise level is left free. In several fits the likelihood maxima lie near the adopted parameter-grid boundaries, and the marginalized intervals are broad and asymmetric. This is particularly evident for the detrended K band, where the intrinsic red-noise component is weak relative to the measurement noise. The anchored broken power-law model avoids this problem by replacing the poorly constrained white-noise parameter with an independently measured quantity, thereby reducing the effective freedom of the background model and providing a clearer physical interpretation.

To evaluate the effect of the background model on the candidate significance, we generated $10^4$ Monte Carlo realizations for each fitted model and computed the local and global FAPs of the candidate GLSP peak. These $10^4$-realization runs are used only for this model-comparison sensitivity test; the fiducial anchored Broken PL (Connolly) significance estimates reported remain based on $10^5$ realizations. 

As shown in Table \ref{tab:psd_background_models}, the residuals and FAPs of the three independently fitted models are broadly comparable in the C band, and they are also consistent with the fiducial Broken PL (Connolly) model used in the main text. For the detrended K band, the anchored Broken PL (Connolly) model yields a slightly larger residual than the freely fitted models, because its high-frequency white-noise level is strictly anchored to the independent measurement-error estimate ($P_{\rm white} = 0.180$ Jy$^2$ d), whereas the free MLE fits tend to underestimate or degenerate the white-noise term in this nearly white-noise-dominated regime. Despite this difference in residual, the resulting K-band FAPs from all models remain at the same modest level (local FAPs $\sim 0.3\%$--$1.4\%$, global FAPs $\sim 1.5\%$--$5.5\%$), confirming that the K-band evidence for the candidate timescale is consistently weak regardless of the background modeling choice.

In summary, the characteristic timescale of the candidate recurrent modulation is stable under the preprocessing and background-model choices tested here. The exact significance, however, is model-dependent, particularly for the K band. These tests do not establish a definitive QPO detection, but they support the conservative interpretation adopted in the main text: the $\sim 1000$ d feature is a candidate recurrent modulation whose statistical significance is conditional on the adopted stochastic background and the limited number of observed cycles.

\section{Markov chain Monte Carlo Fitting Method} \label{appendix: mcmc}

\begin{figure*}[htbp]
    \centering
    \begin{minipage}[b]{0.48\textwidth}
        \centering
        \includegraphics[width=\textwidth]{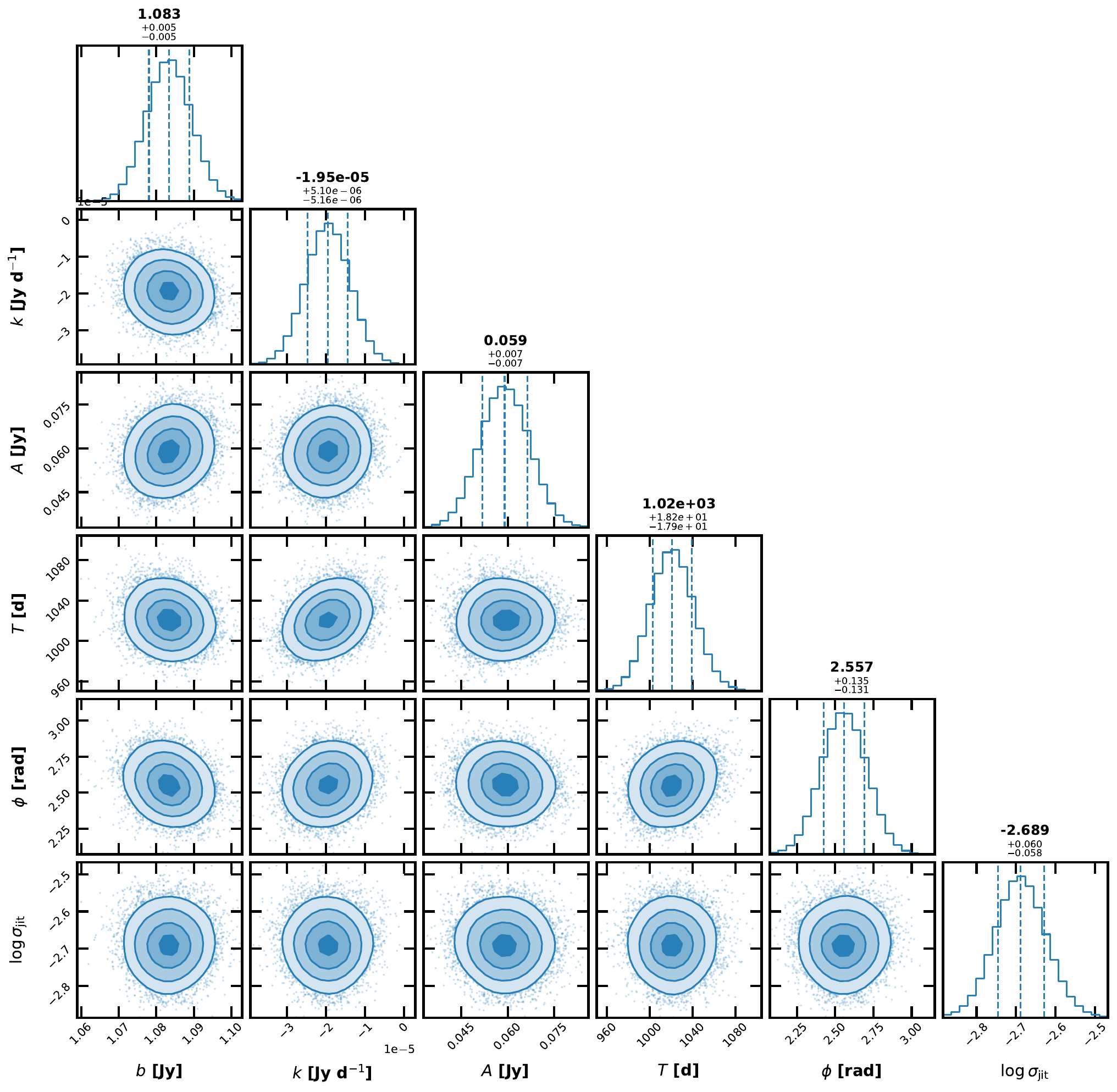}
        \vspace{0.5em} 
        \textbf{(a)}
    \end{minipage}
    \hfill 
    \begin{minipage}[b]{0.48\textwidth}
        \centering
        \includegraphics[width=\textwidth]{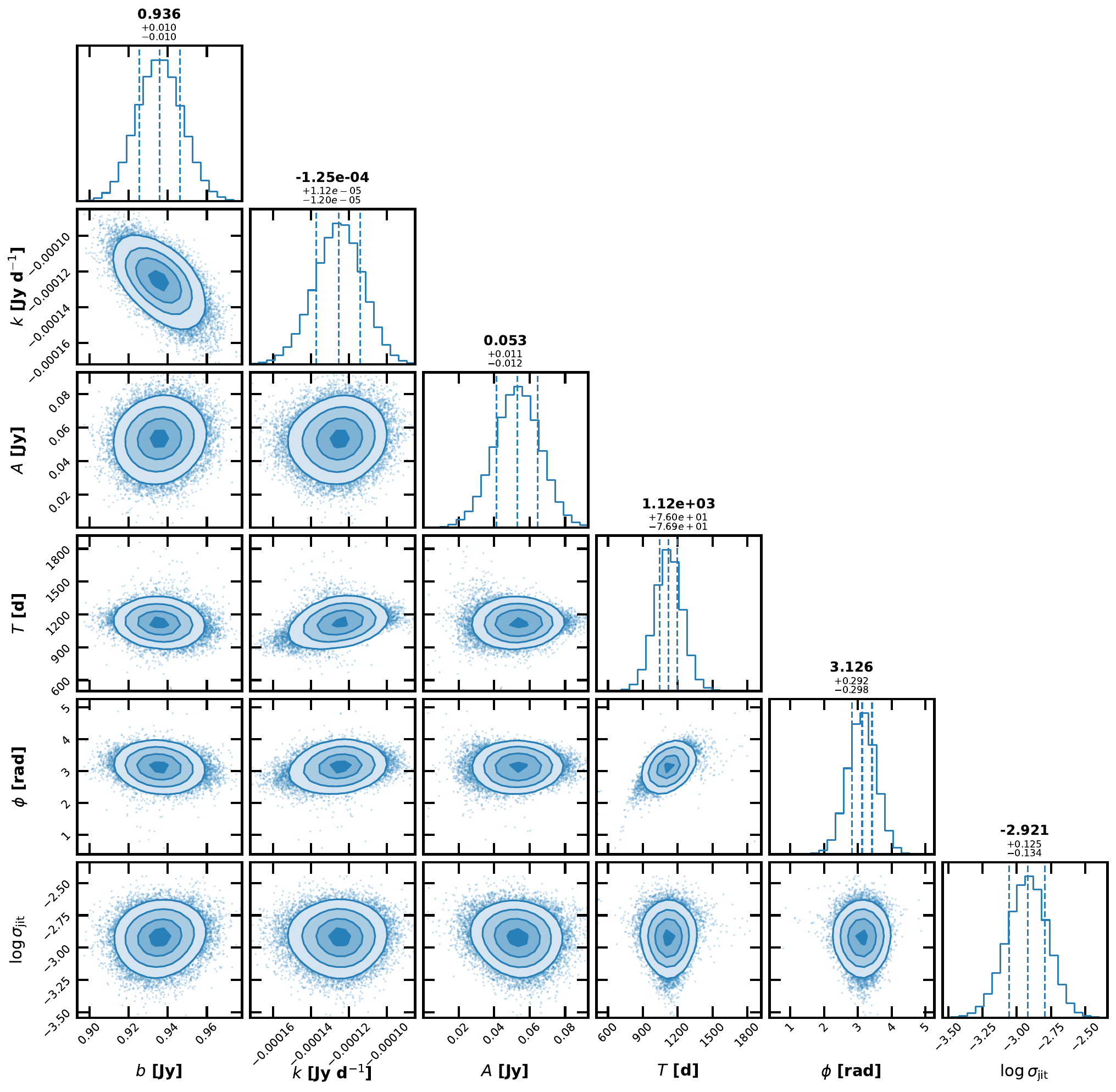}
        \vspace{0.5em}
        \textbf{(b)}
    \end{minipage}
    
    \caption{Parameters corner plot of independent MCMC fitting results.}
    \label{fig:cn_ind}
\end{figure*}

\begin{figure*}
    \centering
    \includegraphics[angle=0,scale=0.28]{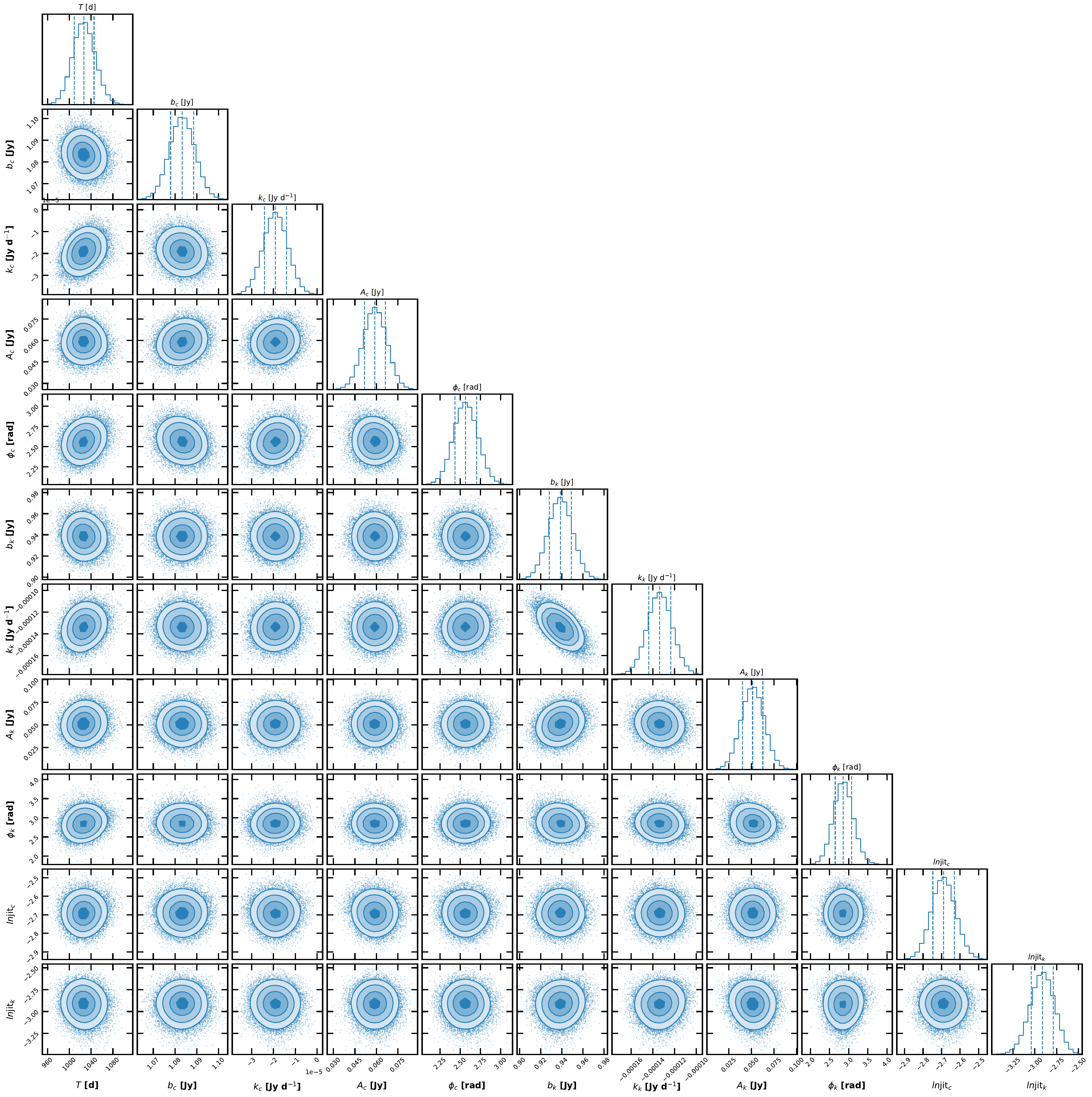}

    \caption{Parameters corner plot of joint MCMC fitting results}
    \label{fig:cn_joint}
\end{figure*}

We used the Markov chain Monte Carlo (MCMC) method to sample the posterior probability distributions of the model parameters. After removing the initial burn-in phase and accounting for the autocorrelation between successive samples, the remaining chain can be used to estimate the posterior distributions and credible intervals of the model parameters \citep{2010CAMCS...5...65G}. In this work, we used the ensemble sampler implemented in the Python package \texttt{emcee} \citep{2013PASP..125..306F}. This method has been widely used in astrophysical parameter inference for its efficiency in exploring correlated parameter spaces.

\paragraph{Model and Likelihood.} The observed light curves are phenomenologically modeled as the combination of a secular linear trend component and a \rev{sinusoidal} component:
\begin{equation}
F(t)=b+k(t-t_0)+A\sin\left({2\pi(t-t_0)}/{T}+\phi\right),
\end{equation}
where $b$ is a flux normalization, $k$ represents the long-term linear evolution rate, $A$ is the modulation amplitude, $T$ is the characteristic period, and $\phi$ is the phase offset. The reference time $t_0$ is adopted as 59400 to reduce parameter covariance.

Assuming independent Gaussian uncertainties, the likelihood function is written as:
\begin{equation}
\ln \mathcal{L}
=
-\frac{1}{2}
\sum_i
\left[
\frac{(F_i-F(t_i))^2}{\sigma_i^2+\sigma_{\rm jit}^2}
+
\ln(2\pi(\sigma_i^2+\sigma_{\rm jit}^2))
\right],
\end{equation}
where $F_i$ and $\sigma_i$ are the observed flux density and measurement uncertainty, respectively. An additional jitter term $\sigma_{\rm jit}=e^{\ln{\rm jit}}$ is introduced to account for possible underestimated observational uncertainties and intrinsic scatter.

\paragraph{Independent and Joint fits.} We first performed independent MCMC fitting for the C- and K-band light curves. For each frequency band, the free parameters were \(\{b,k,A,T,\phi,\ln{\rm jit}\}\), with uniform priors $-10^4< b< 10^4$, $-1< k< 1$, $A\ge0$, $300<T<3000$, $0\le\phi<2\pi$, and $-10<\ln{\rm jit}<10$. The purpose of independent fitting is to investigate whether the two bands exhibit consistent characteristic timescales and different long-term evolution behaviors. The posterior distributions of all parameters were obtained independently for the two bands, and the median values with 16 and 84 percentile intervals are adopted as the parameter distributions.

To test the hypothesis that a single physical modulation drives both bands, we further performed a joint fit that the two bands share the same period $T$ while retaining band-specific $\{b,k,A,\phi,\ln{\rm jit}\}$, i.e., an eleven-parameter model $\{T,b_{\rm C},k_{\rm C},A_{\rm C},\phi_{\rm C},\ln{\rm jit}_{\rm C},b_{\rm K},k_{\rm K}, A_{\rm K},\phi_{\rm K},\ln{\rm jit}_{\rm K}\}$. This shared-period fit is a strong assumption of a common origin for the candidate modulation. In this scenario the likelihood is $\ln\mathcal{L}_{\rm joint}=\ln\mathcal{L}_{\rm C}+\ln\mathcal{L}_{\rm K}$, with the same priors above. We used 64 walkers and ran the chain for up to 300,000 steps. The joint posterior directly provides the inter-band phase difference $\Delta\phi=\phi_{\rm C}-\phi_{\rm K}$ and the corresponding time lag $\Delta t=T({\Delta\phi}/{2\pi})$, defined such that $\Delta t<0$ indicates the K-band modulation leads the C band.

\paragraph{Convergence, burn-in, and thinning.} Convergence was monitored through the integrated autocorrelation time $\tau$ of each parameter chain \citep{2013PASP..125..306F}. A run was terminated when (i) the post-burn-in chain length exceeded $50 \tau$ for every parameter, and (ii) the maximum relative change of $\tau$ between two consecutive checks fell below 1\%, i.e., the autocorrelation times had stabilized. We discarded the first 30\% of each chain as burn-in and thinned the remaining samples by a factor of \({\bf int}[{\bf max}(\tau)]\) to obtain quasi-independent samples . The final effective (quasi-independent) posterior sample sizes are 9184, 26176, and 26368 for the independent C, independent K, and joint fits, respectively, with mean acceptance rates of 0.51.

\paragraph{Results.} The posterior distributions of the model parameters are shown in the corner plots of Figure \ref{fig:cn_ind} and Figure \ref{fig:cn_joint}. The parameter constraints reported in this work correspond to the median values of the posterior distributions, while the uncertainties represent the 16/84 percentiles. All the results are summarized in Table \ref{tab:mcmc}. The independent fits return $T_{\rm C}=1020.8^{+18.2}_{-17.9}$\,d and $T_{\rm K}=1122.0^{+76.0}_{-76.9}$\,d, while the shared-period fit yields $T=1026.7^{+18.7}_{-17.7}$\,d, all consistent with each other and with the GLSP peak periods Section \ref{sec:qpo}. And the joint fit gives phase offset  {$\Delta\phi=-0.288^{+0.242}_{-0.253}$\,rad}, and estimated time lag $\Delta t=(\phi_{\rm C}-\phi_{\rm K})T/2\pi=-47.12_{-41.58}^{+39.57}$\,d (K leading), also support the cross-correlation result in Section \ref{sec:ccf}.

\section{Correlation Analysis And Flux Randomization} \label{appendix:ccf}

All correlation analyses in this work, including the auto-correlation function in Section \ref{sec:qpo} and the cross-correlation function in Section \ref{sec:ccf}, are computed with the z-transformed discrete correlation function (ZDCF) method \citep{1997ASSL..218..163A}, which is well suited for unevenly sampled time series. For the ACF, we adopt the first positive correlation peak at non-zero lag as the characteristic timescale; the lag of this peak and its positional uncertainty, taken as the extent of the corresponding ZDCF lag bin, are reported as the ACF estimation result in Table \ref{tab:qpo_results}.

For the CCF, the correlation peaks are broad and not sharply peaked, so a single-point peak lag is not a robust statistic. \rev{Thus, we estimate the centroid lag of the central correlation hump using flux randomization. For each case (raw C-K and detrended C-K) we generate 100,000 realizations; at the fixed observed epochs, the flux densities are drawn from Gaussian distributions centered on the observed values with standard deviations equal to the measurement uncertainties for each epoch.} The cross-ZDCF of each realization is computed with the same binning parameters (${\rm minpts}=50$) as adopted for the observed data.

For each realization we first locate the two local minima of the cross-ZDCF near lags of $\pm500$\,d, which correspond to the first anti-correlation dips of the $\sim1000$\,d quasi-periodic modulation, and define the central hump as the contiguous region between them. After subtracting a tilted baseline connecting the two minima, we compute the centroid lag as the width- and height-weighted mean, $\tau_{\rm cent}=\sum_i w_i h_i \tau_i/\sum_i w_i h_i$, over the bins whose baseline-subtracted correlation exceeds 50\% of the hump maximum. Realizations whose hump height falls below 70\% of the hump height measured on the observed data are rejected, as their correlation structures are too flat to provide a meaningful centroid. The accepted centroids form the lag distribution $p(\tau\,|\,D)$ for each case, whose medians and 16/84 percentiles are reported in Section~\ref{sec:ccf}.


\bibliography{sample701}{}
\bibliographystyle{aasjournalv7}

\end{CJK*}
\end{document}